\documentclass{IEEEtran}
\usepackage[utf8]{inputenc}
\usepackage{microtype}

\usepackage{graphicx}
\usepackage{epstopdf}
\usepackage{amsmath,amssymb,booktabs,tabularx}
\usepackage{bm}
\usepackage[hyphens]{url}
\usepackage{hyperref}
\hypersetup{
    colorlinks,
    linkcolor={red!75!black},
    citecolor={blue!75!black},
    urlcolor={blue!75!black}
}
\usepackage[capitalise]{cleveref}

\usepackage{mathrsfs}
\usepackage{float}
\usepackage{multirow}
\usepackage{soul}
\usepackage[table]{xcolor}
\usepackage{comment}
\usepackage{subcaption}
\usepackage{mwe}
\usepackage{slashbox}
\usepackage{authblk}
\usepackage{multicol}
\usepackage{import}
\usepackage{siunitx}
\usepackage{mathtools}
\usepackage{array}
\usepackage{tabularx}

\usepackage{makecell} 

\makeatletter
\DeclareRobustCommand\onedot{\futurelet\@let@token\@onedot}
\newcommand{\@onedot}{\ifx\@let@token.\else.\null\fi\xspace}
\makeatother

\newcommand{\cf}{cf\onedot}

\newcommand{\etal}[1]{#1 \emph{et~al\onedot}}
\newcommand{\ie}{i.\,e.,\xspace}

\makeatletter
\let\old@ps@headings\ps@headings
\let\old@ps@IEEEtitlepagestyle\ps@IEEEtitlepagestyle
\def\confheader#1{%
	\def\ps@IEEEtitlepagestyle{%
		\old@ps@IEEEtitlepagestyle%
		\def\@oddhead{\strut\hfill#1\hfill\strut}%
		\def\@evenhead{\strut\hfill#1\hfill\strut}%
	}%
	\ps@headings%
}
\makeatother

\confheader{%
	\parbox{20cm}{SUBMITTED TO IEEE TRANSACTIONS ON GEOSCIENCE AND REMOTE SENSING FOR POSSIBLE PUBLICATION.}
}

\title{On Mathews Correlation Coefficient and Improved Distance Map Loss for Automatic Glacier Calving Front Segmentation in SAR Imagery}

\author{Amirabbas Davari*, Saahil Islam*, Thorsten Seehaus, Matthias Braun, Andreas Maier, Vincent Christlein
\thanks{* AmirAbbas Davari and Saahil Islam contributed equally to this work.}
\thanks{A.\ Davari, S.\ Islam, A.\ Maier and V.\ Christlein are with the Computer Science department at Friedrich-Alexander University Erlangen-Nürnberg, 91058 Erlangen, Germany (email: amir.davari@fau.de).}
\thanks{T.\ Seehaus and M.\ Braun are with the Geography \& Geosciences department at Friedrich-Alexander University Erlangen-Nürnberg, 91058 Erlangen, Germany.}}

\begin{document}
\maketitle
\begin{abstract}
The vast majority of the outlet glaciers and ice streams of the polar ice sheets end in the ocean. 
Ice mass loss via calving of the glaciers into the ocean has increased over the last few decades. 
Information on the temporal variability of the calving front position provides fundamental information on the state of the glacier and ice stream, which can be exploited as calibration and validation data to enhance ice dynamics modeling.
To identify the calving front position automatically, deep neural network-based semantic segmentation pipelines can be used to delineate the acquired SAR imagery.
However, the extreme class imbalance is highly challenging for the accurate calving front segmentation in these images. 
Therefore, we propose the use of the Mathews correlation coefficient (MCC) as an early stopping criterion because of its symmetrical properties and its invariance towards class imbalance. 
Moreover, we propose an improvement to the distance map-based binary cross-entropy (BCE) loss function. 
The distance map adds context to the loss function about the important regions for segmentation and helps accounting for the imbalanced data.  
Using Mathews correlation coefficient as early stopping demonstrates an average \SI{15}{\percent} dice coefficient improvement compared to the commonly used BCE. 
The modified distance map loss further improves the segmentation performance by another \SI{2}{\percent}. 
These results are encouraging as they support the effectiveness of the proposed methods for segmentation problems suffering from extreme class imbalances.

\end{abstract}

\begin{IEEEkeywords}
semantic segmentation, class imbalance, Mathews correlation coefficient (MCC), improved distance map loss
\end{IEEEkeywords}

\section{Introduction}\label{sec:introduction}
The mass loss of the Polar ice sheets and other calving glaciers has accelerated in the last two decades. This is one of the main causes of global sea level rise. Currently, the sea level contribution from thermal expansion is exceeded by the contribution from ice sheets, glaciers, and ice caps~\cite{van2009partitioning}.
The retreat of the glacier fronts can produce a re-enforcing effect of ice loss. For example, the retreat from the pinning points can reduce the lateral buttressing forces, which further destabilize the glaciers' ice discharge into the ocean~\cite{furst2016safety}. A significant frontal retreat can further cause a retreat of the grounding zone leading to further destabilization of the glacier flow~\cite{friedl2018recent}. This makes the calving front, the region where the glacial ice breaks off into the ocean. Hence, the calving front position (CFP) is an important variable for monitoring, but also ice dynamic modeling.

The detection of the calving front locations is commonly carried out using optical imagery or Synthetic Aperture Radar (SAR) imagery. Cloud covers and polar nights affect the optical imagery and the data coverage is strongly curbed. On the other hand, SAR acquisitions are unaffected by such situations and can be carried out all year round, making SAR imagery more advantageous for monitoring purposes in polar regions. The common approach to detect CFPs has been visual analysis and manual delineation on the obtained SAR images. However, manual digitization is tedious, subjective, expensive, and time-consuming. 
Moreover, sea ice and calved-off icebergs in the oceans make it often difficult to reliably differentiate between ice mass and these chunks of floating ice, making manual delineation subjective. This leads to a lack of temporally updated data on the position of the calving front, which is considered an essential climate variable (ECV). Automatic detection of CFPs would close this monitoring gap. 
In the past, various algorithms of automatic and semi-automatic detection of the calving front positions had been proposed. Edge detection algorithms such as Roberts edge extractor has been applied to ERS-1 SAR images by \etal{Sohn}~\cite{sohn1999mapping}. \etal{Seale}~\cite{seale2011ocean} have exploited Sobel filter and brightness profiling to detect the CFPs on daily MODIS images. A detailed overview of various automatic and semi-automatic approaches on this application is provided by \etal{Baumhoer}~\cite{baumhoer2019automated}.

Deep learning-based methods such as Convolutional Neural Networks (CNNs)~\cite{minaee2020image} are a better alternative to the traditional automatic delineation of CFPs. Over the past two decades, CNNs have been seen to outperform the above-mentioned manually designed filter-based methods and many such traditional computer vision techniques in the field of image processing~\cite{o2019deep}. The CNN architecture (the type and number of layers and their order) is chosen depending on the application and the training data. The parameters in a CNN are fine-tuned based on optimizing a predefined objective function, also referred to as cost function or loss function. 

Deep convolutional networks can be used on SAR images to automatically classify the image pixels, belonging to semantically different regions, into distinct classes. Such a task is referred to as semantic segmentation. Architectures such as SegNet~\cite{badrinarayanan2017segnet} and U-Net~\cite{ronneberger2015u} are commonly used for semantic segmentation applications~\cite{minaee2020image}. The use of the U-Net architecture for SAR image segmentation was first exploited by \etal{Mohajerani}~\cite{mohajerani2019detection} on Landsat-5, 7, and 8 images to automatically detect calving fronts. \etal{Zhang}~\cite{zhang2019automatically} have used U-Net on higher resolution TerraSAR-X images of Jakobshavn Isbræ glaciers to separate the whole ice mélange regions from non-ice mélange regions, which are subsequently post-processed to manually delineate the calving fronts.

Despite being powerful learning tools, the high number of parameters in Deep Neural Networks (DNNs) comes with the risk of
over-fitting to the training data~\cite{srivastava2014dropout}, \ie it ``memorizes'' the non-predictive features of the training data instead of ``learning'' its underlying distribution
In this paper, one contribution focuses on proper exploitation of early stopping to prevent the network from over-fitting~\cite{overfit}. 

Early stopping suggests stopping training based on the error on independent validation data to maintain high performance on new unseen data from the same distribution. Since the training gets terminated based on the error on the validation data, it is crucial to have the right performance metric as a stopping criterion. The traditional method of early stopping comprises monitoring the loss~\cite{prechelt1998early}. However, the significance of the performance metric increases in cases of class-imbalanced data~\cite{luque2019impact}. An imbalanced dataset comprises classes with an unequal number of class elements. The majority of performance metrics are challenged in the presence of severely imbalanced data. For example, accuracy is not an appropriate metric in applications with class-imbalance as incorrect classifications of the minority class could be overlooked. On the other hand, other metrics such as f1-score greatly diminishes the effect of class imbalance. Mathews Correlation Coefficient (MCC) is theoretically argued to have a higher tolerance over class imbalance than f1-score and confusion entropy~\cite{jurman2012comparison,chicco2020advantages} and was for example used to compare between a vast range of classifiers for evaluation of models on micro-array gene expression and genotyping data~\cite{shi2010maqc}.
In this paper, we are suggesting (a) the use of early stopping to prevent over-fitting and (b) using MCC instead of the classical BCE loss as stopping criterion.

While early stopping handles over-fitting with the right performance metric, it is also crucial to objectively handle the learning of the network on the imbalanced dataset. The imbalanced dataset introduces a bias towards the majority class during the training of the network while the minority class is the point of interest in many real-world problems. Common solutions are for example data re-sampling~\cite{davari2019fast}, cost-sensitive learning, one class learning, or ensemble learning methods~\cite{classimbalance}. While these methods could be effective, they do not guarantee to be so in all kinds of tasks. For example, over-sampling undergoes a risk of early over-fitting because the network uses many samples more than once while under-sampling reduces the size of the training data. Artificially balancing the class distributions was not very successful in some tasks~\cite{standford} and many ensemble-based methods are resource-demanding and time-consuming~\cite{classimbalance}.

Another way of cost-sensitive learning is to weigh the minority class higher to punish miss-classification. In imbalanced segmentation tasks, the loss function can also be trained to penalize more the contour of the segmentation masks than the regions. \etal{Kervadec}~\cite{kervadec2019boundary} used such a boundary-based loss function for medical image segmentation. They argued that region integrals of the segmentation masks highly vary over the classes and instead trained on the space of contours. Another distance-based loss approach was stated to segment knee bones in 3d MRI~\cite{caliva2019distance}. However, they employed the idea of a distance map-based loss because most errors in segmentation are found to be at the edges of the segmented masks. A distance map-based loss was also used by \etal{Adhikari}~\cite{adhikari2020distance} to detect forest trails in optical images of forests, which cover only small regions in the images. Therefore, the trails were weighted higher in the loss function while the other regions were weighted less based on the distance to the forest trails. 

In this paper, we show that the use of MCC as an early stopping criterion is superior to a binary cross-entropy (BCE) monitored network. The performance of the network is further enhanced by an improved distance map-based BCE loss and is compared with (a) the traditional BCE loss and (b) a weighted BCE loss. The rest of this paper is organized as follows: \cref{sec:background} reviews the tools that have been used in our contributions, \ie Mathews correlation coefficient, and the distance map loss. \cref{methodology} describes our proposed pipeline and explains the main contributions of this paper in detail. Next, the dataset description, experimental setup, and the quantitative and qualitative results are presented and discussed in \cref{sec:experimental_setup}. Finally, \cref{sec:conclusion} concludes the paper.

\section{Theoretical Background}\label{sec:background}

\subsection{Mathews Correlation Coefficient}\label{MCC}  

Binary cross-entropy is a popular loss function in binary classification scenarios. A good loss is usually supposed not to produce very small or extremely large gradients, which is one of the desirable properties of the BCE loss~\cite{Goodfellow-et-al-2016}. We could choose to use the validation BCE loss also as an early stopping criterion. However, BCE loss is very fragile in the presence of class-imbalance~\cite{jadon2020survey}. Hence, it becomes an unfavorable early stopping criterion for class-imbalanced segmentation.

On the other hand, MCC represents true positives (TP), false negatives (FN), false positives (FP) and true negatives (TN) in a balanced manner~\cite{baldi2000assessing}. It is balanced in the sense that the rate of positive class predictions and negative class predictions are given equal importance. MCC is defined as:
\begin{equation}\label{eq:mcc}
	\text{MCC}=\frac{\text{TP} \times \text{TN} - \text{FP} \times \text{FN}}{\sqrt{(\text{TP}+\text{FP})(\text{TP}+\text{FN})(\text{TN}+\text{FP})(\text{TN}+\text{FN})}}\enspace.
\end{equation}
It lies in the range $[-1,1]$ with $1$ representing the perfect prediction, $-1$ denoting wrong, and $0$ suggesting random prediction. MCC takes positives and negatives equally into account and provides very little ``discrimination'' to random acts~\cite{jurman2012comparison}. 

\subsection{Distance Map Loss}\label{DMap}
While BCE is the most commonly used loss function in most deep learning-based methods, it is not favorable for class-imbalance problems. A common approach to tackle the class-imbalance with BCE loss is to weigh each class with its inverse class frequency. However, it does not always guarantee to produce better performance~\cite{standford}. Moreover, minority classes with higher weights can lead to noisy gradients~\cite{adhikari2020distance}. On the other hand, a distance map-based approach adds context to the loss functions based on the regions of importance in the segmentation task. Distance maps, also known as distance fields or distance transforms, are derived representations of an image based on a distance metric. In computer graphics, distance maps are often used for volume rendering to generate offset surfaces or a blending between surface models~\cite{frisken2000adaptively}. 

A distance map-based loss trains the network by penalizing the predictions with respect to its distance from the object of interest in the image. 
A distance map-based loss is also beneficial for class imbalance. By forcing the network to give more attention to the regions of interest (minority class), it counters the bias of the network's training towards the majority class. The distance map loss works on the idea of weighting far away pixels (from the semantic line) higher than the ones in its proximity~\cite{adhikari2020distance}. The loss is formulated as:
\begin{equation}\label{eq:ldmap}
	\begin{split}
		J_\text{DW} &= \frac{1}{M \cdot N}\sum\limits_{j=1}^{N}\sum\limits_{i=1}^{M} ((1 - g_{i,j}) \cdot d_{i,j} \cdot p_{i,j} \\
		&- g_{i,j} \cdot d_\text{max} \cdot \log(p_{i,j})) \enspace,
	\end{split}
\end{equation}
with $g$ being the ground truth map and $p$ the prediction at pixel location $i,j$. The distance map is denoted as $d$, with $d_\text{max}$ being the maximum distance, and $M,N$ represent the spatial dimensions of the maps.

\section{Methodology}\label{methodology}
\subsection{Processing Pipeline}
\begin{figure}[t]
		\includegraphics[width=1\linewidth]{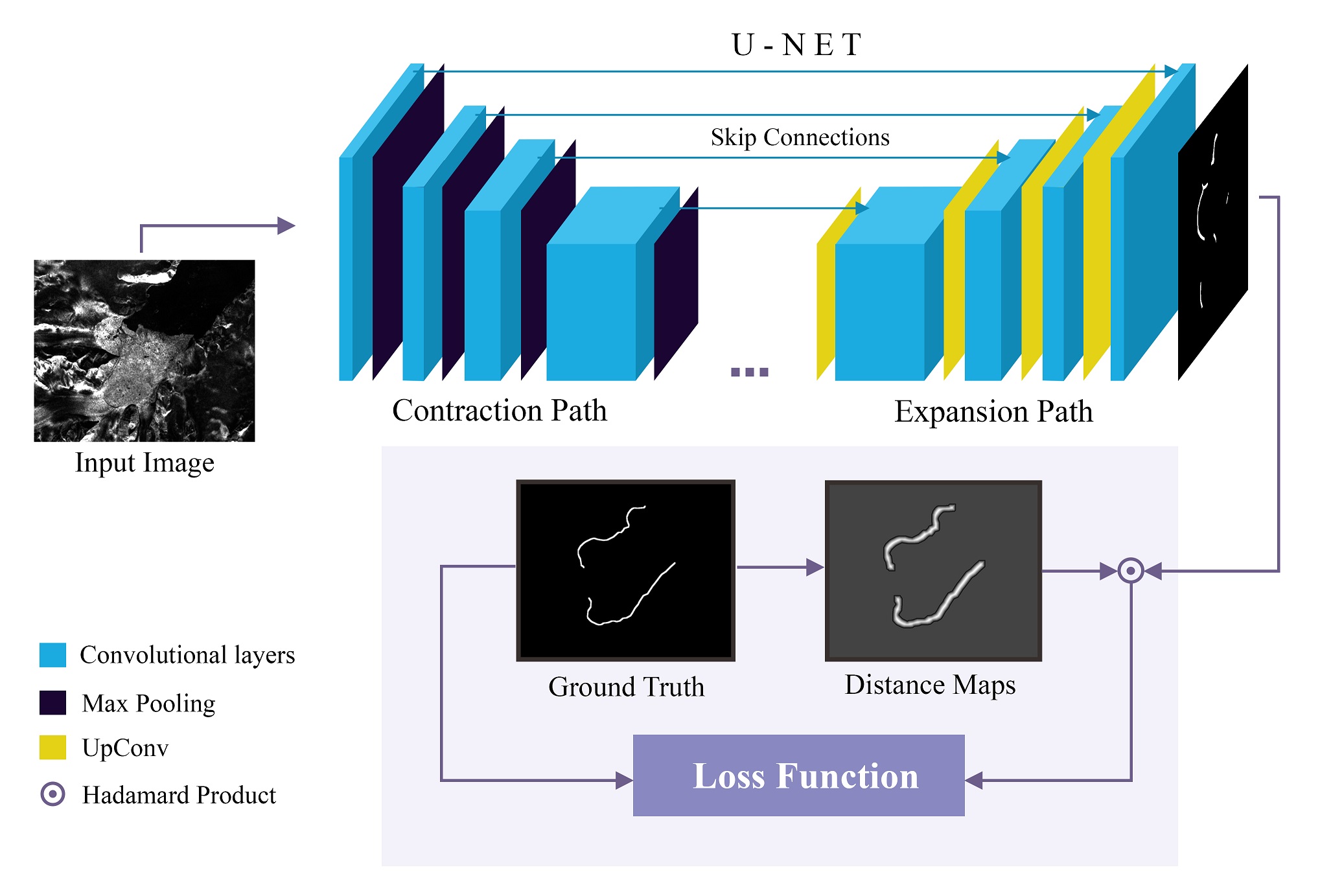}
	%
	\caption{Proposed glacier front segmentation pipeline.}
	\label{fig:dmapLoss}
\end{figure}
\Cref{fig:dmapLoss} outlines our proposed segmentation workflow. 
We perform the segmentation of calving fronts images using a U-Net architecture. The SAR images are used as input to the network, which learns either to predict the zones of ice mass regions or the calving front lines, depending on the ground truth data presented to the model at training time. In the rest of this paper, we refer to the ice/non-ice regions as \emph{zones} and calving front locations as \emph{front lines} for convenience. In the variant that the network is trained to predict the ice zones, the predictions are post-processed to extract the boundaries and in this way obtain the front lines.

The images are first pre-processed to remove noise and then patches of a predefined size were extracted from both the images and their corresponding ground truth before they are fed to the network. Before extracting the patches, the images are padded with zeros to the next larger image size that is a multiple of the patch size. The predicted masks on the patches are later stitched together to reconstruct the original size image, and the padding is discarded.

\subsection{MCC as the Early Stopping Metric}
The data is split into three subsets, namely training, validation, and test set. We use early stopping to stop the training before it overfits. However, the validation error is usually very noisy. Hence, to ensure that the training terminates where the validation performance is optimum, we pre-define a patience number. This is the number of epochs the early stopper waits to check for improvement on the validation error before terminating the training when no improvement is noticed. The common practice is to monitor the BCE loss on the validation set. However, due to the severe class-imbalance, we propose the use of the Mathews correlation coefficient (MCC) as an early stopping metric as it is theoretically more robust towards class-imbalance.

\subsection{Modified Distance Map and Distance Map BCE Loss}\label{mod_DMap}
\begin{figure}[tbhp]
	\begin{minipage}[b]{0.32\linewidth}
		\centering
		\centerline{\includegraphics[width=1\linewidth]{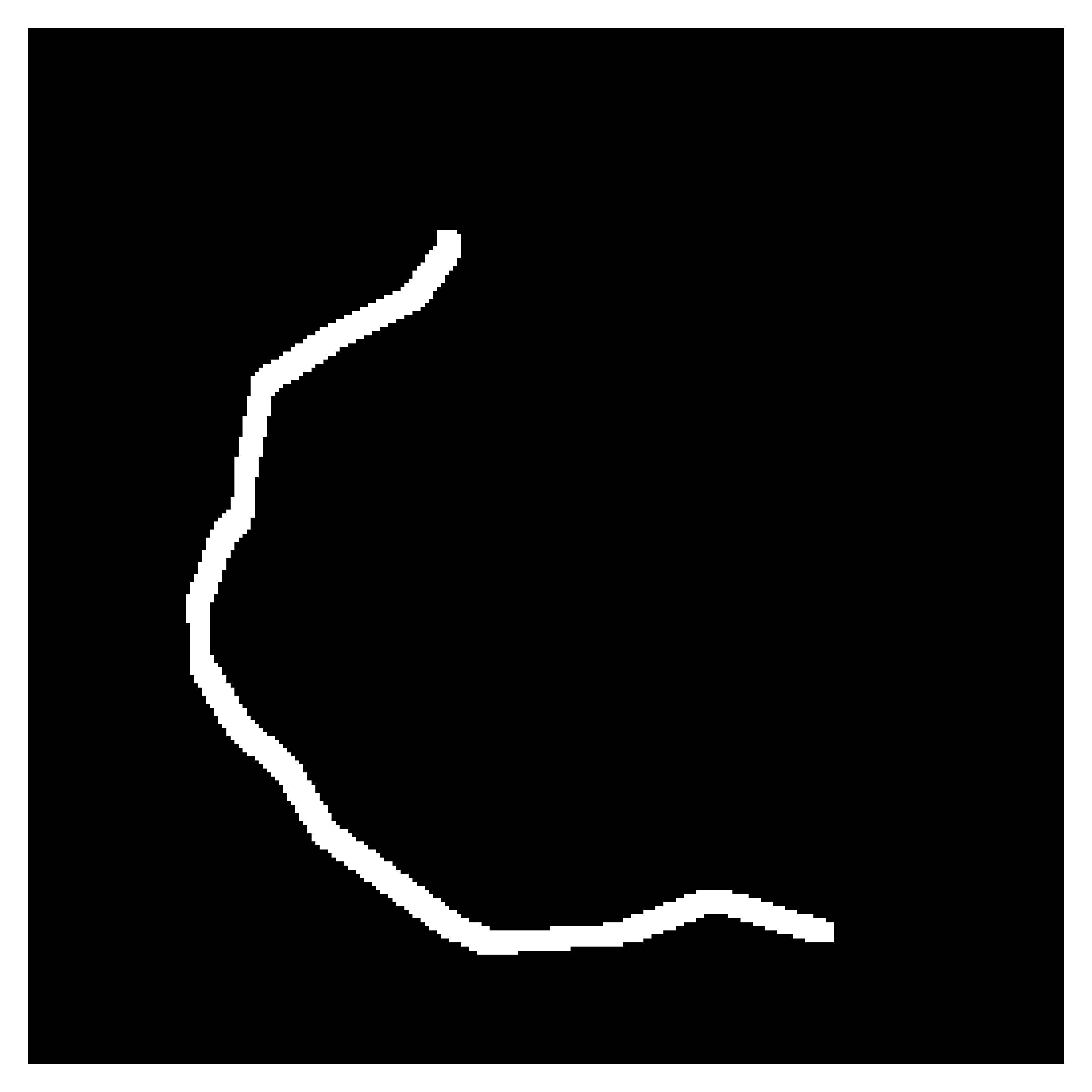}}
		\centerline{(a)}\medskip
	\end{minipage}
	\begin{minipage}[b]{0.32\linewidth}
		\centering
		\centerline{\includegraphics[width=1\linewidth]{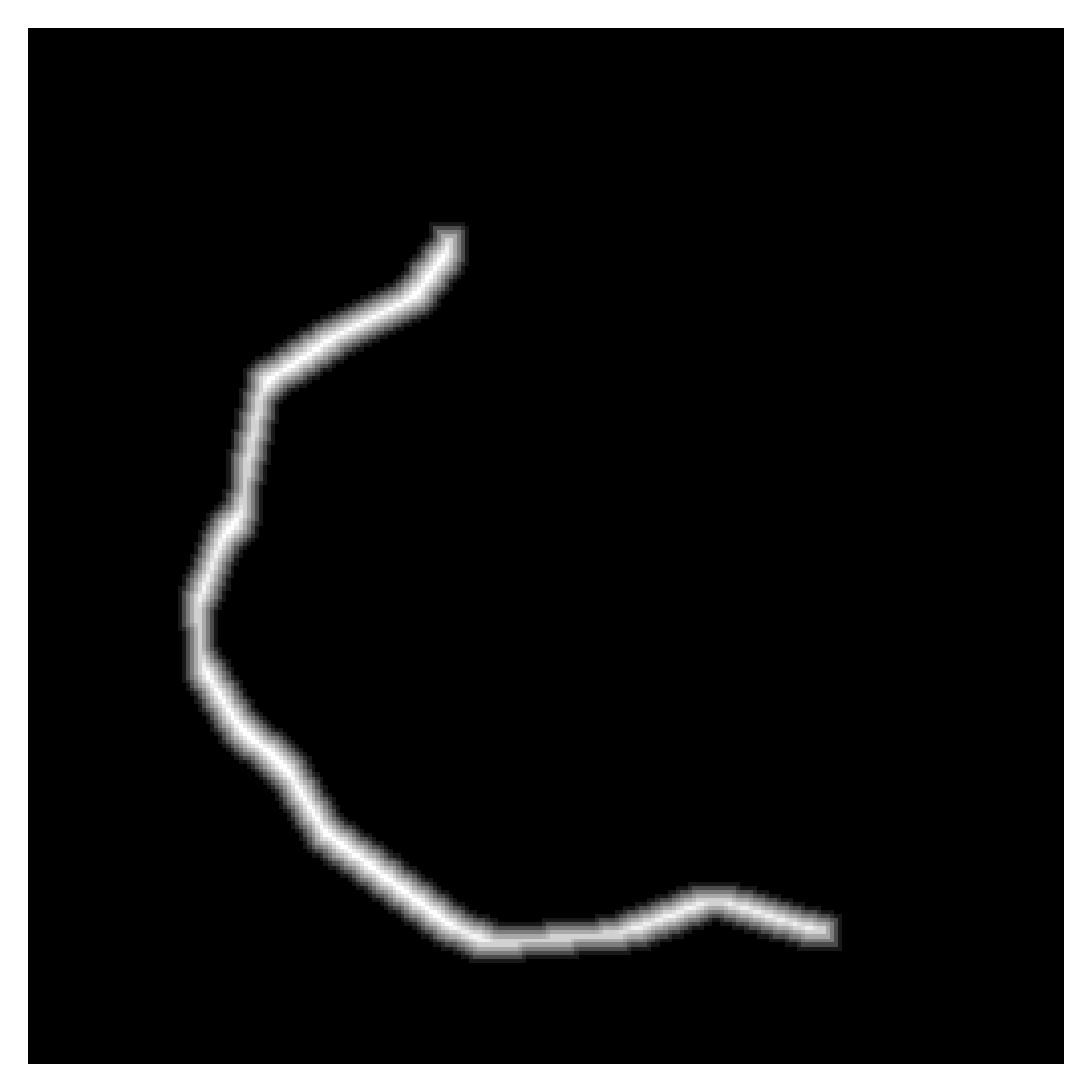}}
		\centerline{(b)}\medskip
	\end{minipage}
	\begin{minipage}[b]{0.32\linewidth}
		\centering
		\centerline{\includegraphics[width=1\linewidth]{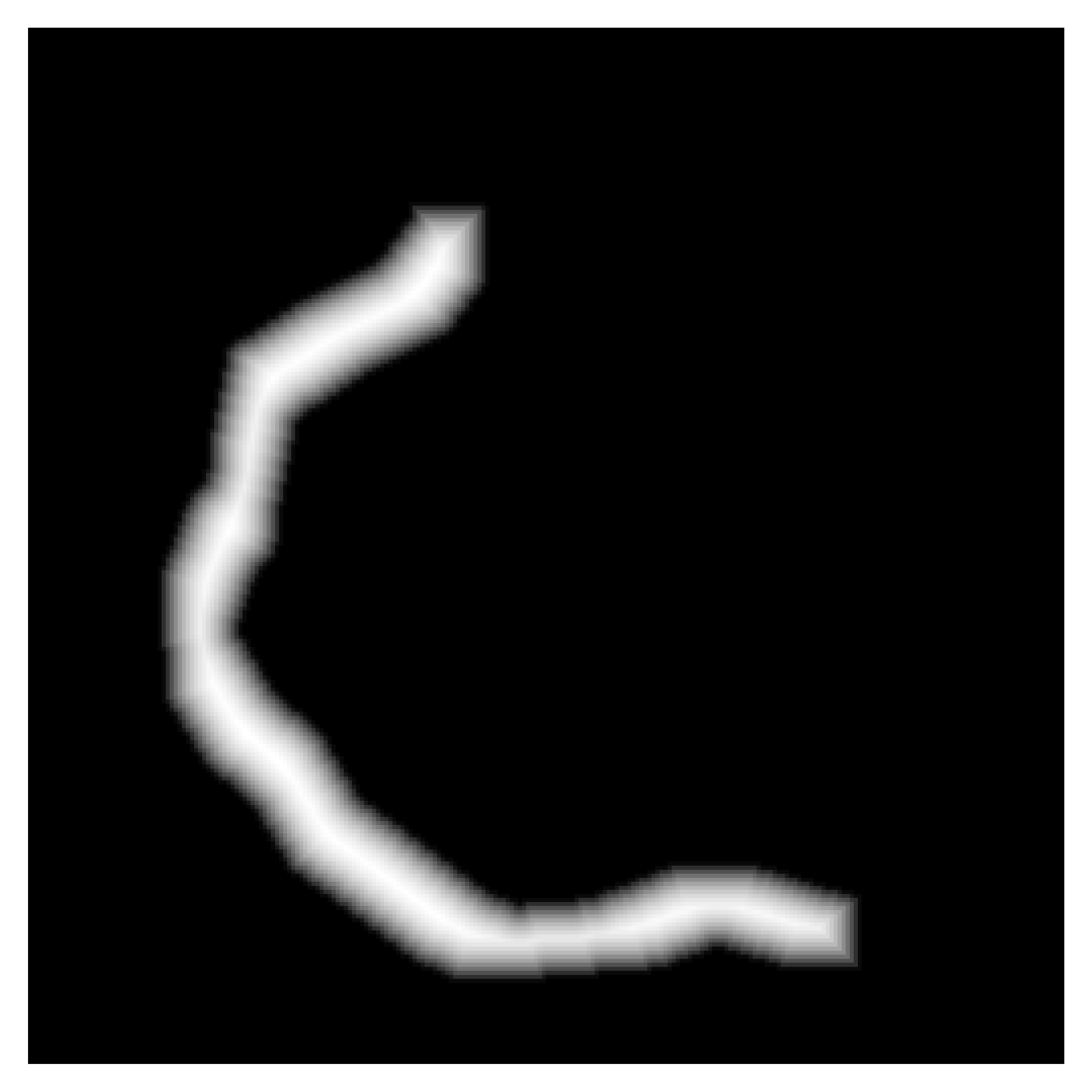}}
		\centerline{(c)}\medskip
	\end{minipage}
	\begin{minipage}[b]{0.32\linewidth}
		\centering
		\centerline{\includegraphics[width=1\linewidth]{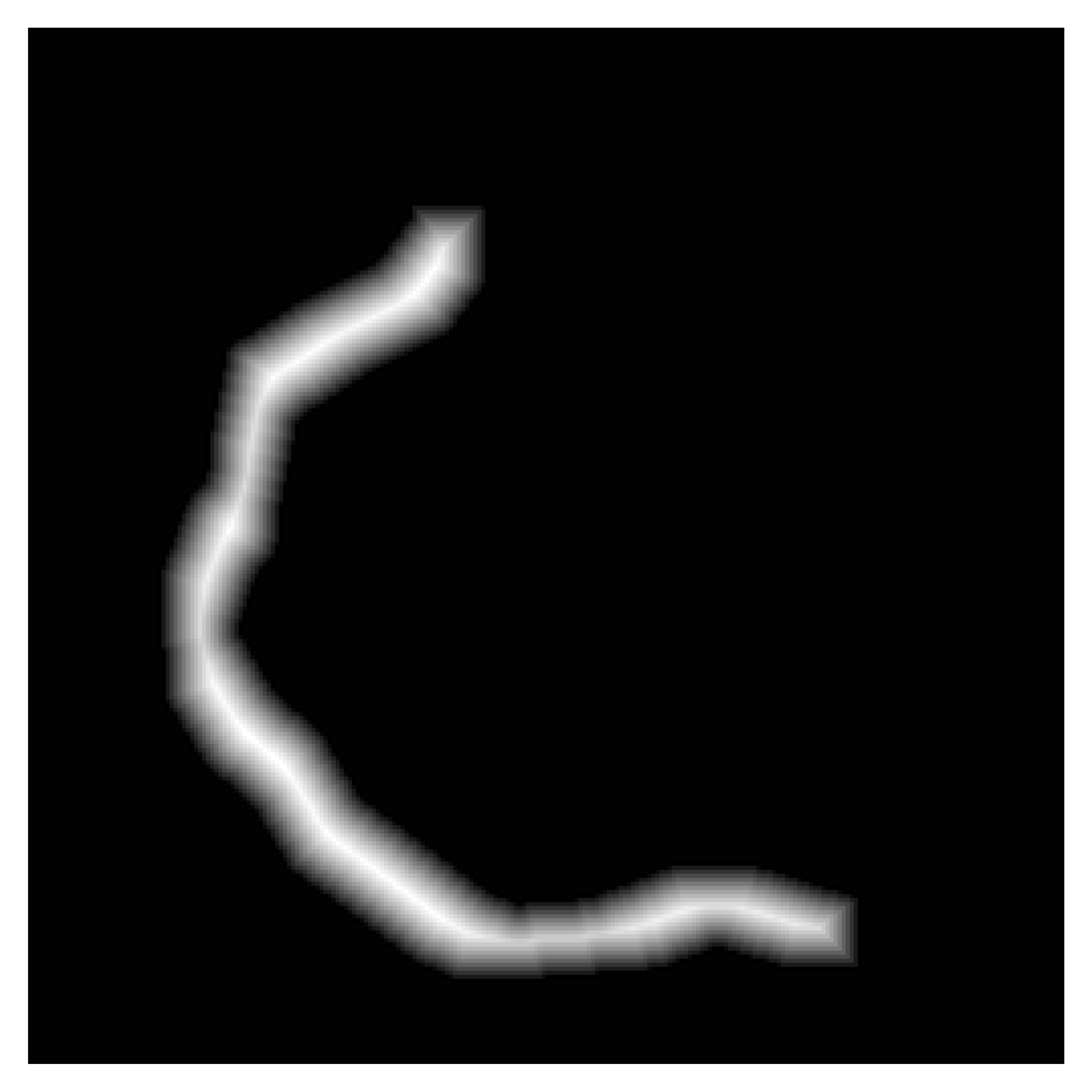}}
		\centerline{(d)}\medskip
	\end{minipage}
	\begin{minipage}[b]{0.32\linewidth}
		\centering
		\centerline{\includegraphics[width=1\linewidth]{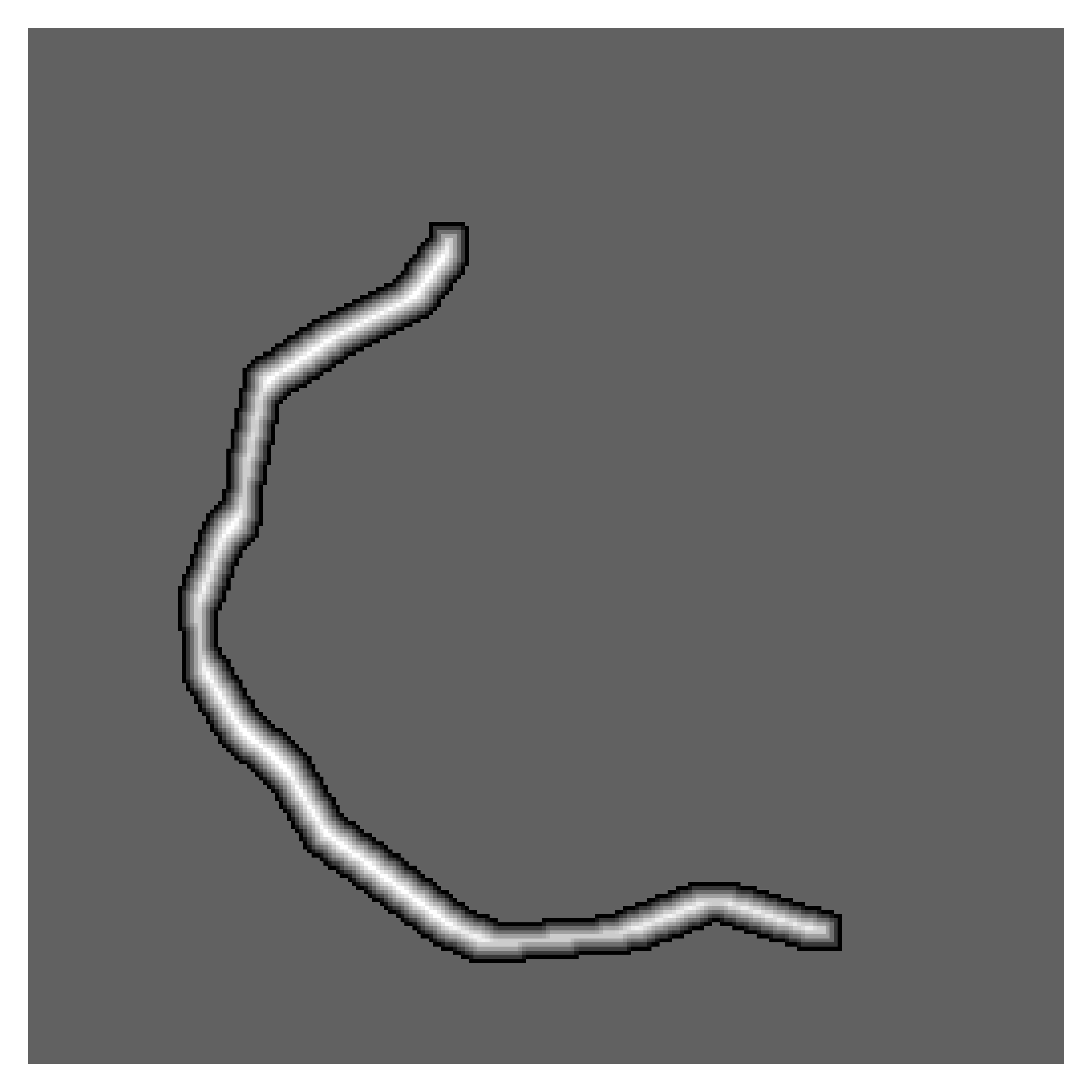}}
		\centerline{(e)}\medskip
	\end{minipage}
	\begin{minipage}[b]{0.32\linewidth}
		\centering
		\centerline{\includegraphics[width=1\linewidth]{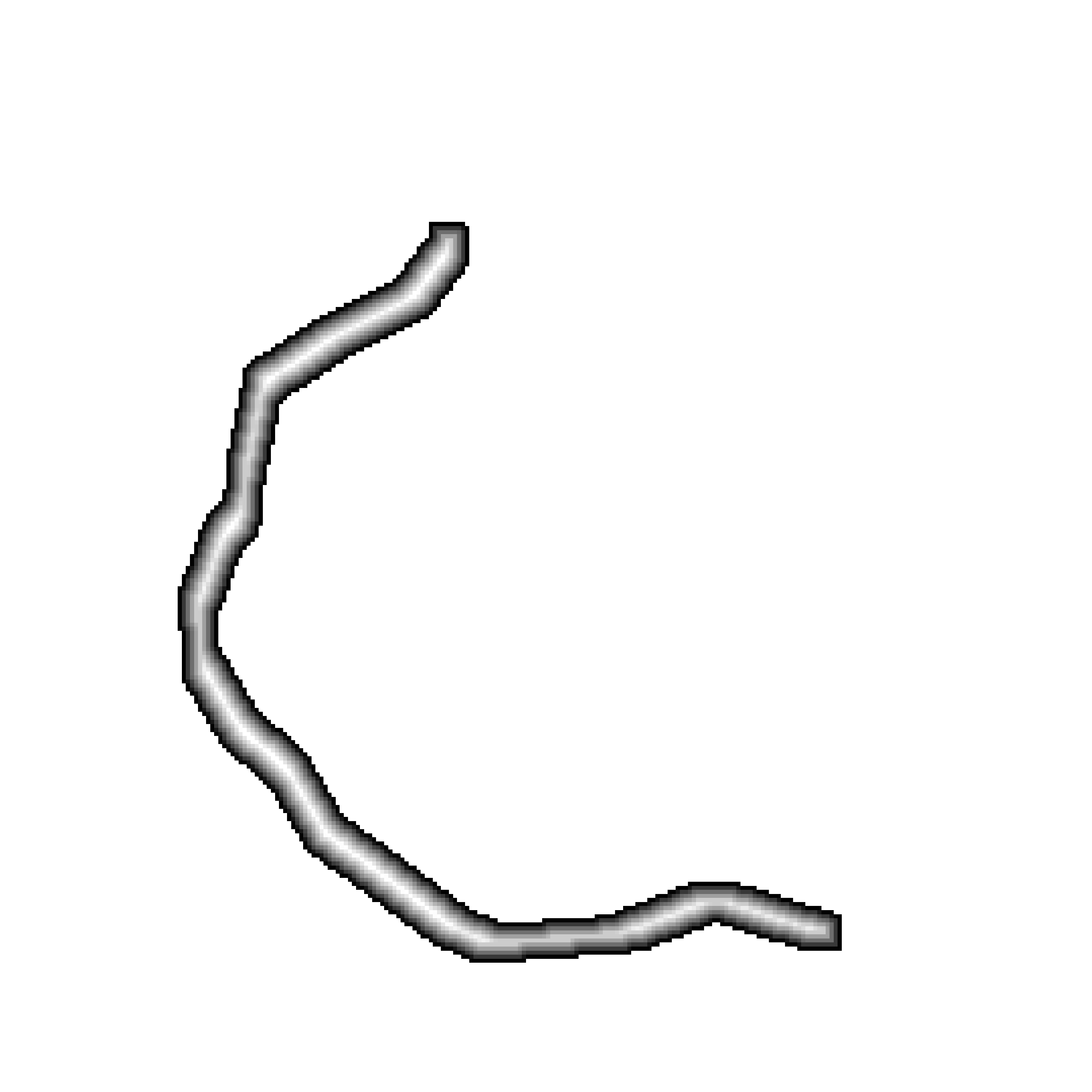}}
		\centerline{(f)}\medskip
	\end{minipage}
	\caption{Variation of the distance maps with different parameters when applied on the ground truth. First image is (a) ground truth, and rest are distance maps with (b) $w=5$, $R=4$, $k=0.1$, (c) $w=15$, $R=4$, $k=0.1$, (d) $w=15$, $R=8$, $k=0.1$, (e) $w=5$, $R=4$, $k=0.5$, and (f) $w=5$, $R=4$, $k=1$.}
	\label{fig:distmap}
\end{figure}

We formulate the distance map in a generalized form to be used in different loss functions. The computed distance map weighs the line locations higher while the regions around the proximity of the front lines are weighed less based on the distance to the lines. As we move progressively away from the front line locations, the weights keep reducing until they become $0$. In other words, true line locations are given high importance and the proximity adds relaxation to the network. The points beyond which these weights become zero are considered to be the background (majority class). 
In contrast to the distance map-based approach by \etal{Adhikari}~\cite{adhikari2020distance}, \cf \cref{eq:ldmap}, we weigh the background differently depending on the class imbalance. In particular, we calculate the distance map from the ground truth masks as follows:

We first thicken the ground truth lines by morphological dilation ($\delta_w$)~\cite{soille2003morphological}. Then, we use the Euclidean distance transform (EDT), divided by a constant factor of $R$. The strength of relaxation is determined by $R$, \ie how smoothly the weights fall off from the front lines to the background. The morphological dilation of the ground truth images thickens the front lines based on the structuring element used. We use a rectangular structuring element with a size of $(w,w)$, where $w$ determines the amount of thickening of the lines. 
It is further passed through a sigmoid function $\sigma$ to obtain values in the range of $[0,1]$:
\begin{equation}\label{eq:dilate}
	y_\text{dl} = \delta_{w}(y)\enspace,
\end{equation}   
\begin{equation}\label{eq:edt}
	y_\text{edt} = \sigma\left(\frac{\text{EDT}(y_\text{dl})}{R}\right)\enspace,
\end{equation}
where $y$ denotes the ground truth. Finally, the distance map $d$ is obtained by:
\begin{equation}\label{eq:dmap}
	d = y_\text{edt} + k(1 - y_\text{dl})\enspace,
\end{equation}
where $k$ is a constant in the range of $(0,1]$ determining the weight given to the background, with $k=1$ signifying utmost importance. Since we deal with high class imbalance, a high weight is not desirable. However, it is a hyper-parameter that depends on the dataset and its class imbalance ratio, and needs to be tuned. Various distance maps are depicted with different $w$, $R$, and $k$ in \cref{fig:distmap}. 
Finally, the distance map computed on the ground truth is used in the BCE loss to create the distance map BCE loss. The BCE loss ($\text{J}_{\text{BCE}}$) is given by:
\begin{equation}\label{eq:bce}
	\text{J}_{\text{BCE}}(y,\hat{y}) = -\frac{1}{N}\sum\limits_{i=1}^{N}(y_i\log \hat{y}_i + (1-y_i)\log (1-\hat{y}_i))\enspace,
\end{equation}
where $\hat{y}$ is the prediction, $y$ is the ground truth, and $N$ is the batch size during training.
Incorporating the distance map into the BCE loss ($\text{J}_{\text{DBCE}}$) results in:
\begin{equation}
	\begin{split}
		\text{J}_{\text{DBCE}}(y,\hat{y}) &= -\frac{1}{N}\sum\limits_{i=1}^{N}(y_i\log(\hat{y}_i \cdot d_i) \\
		&+ (1-y_i)\log (1-\hat{y}_i \cdot d_i)\enspace.
	\end{split}
\end{equation}
%

\section{Experimental Setup and Evaluation}\label{sec:experimental_setup}
\subsection{Dataset}
\begin{figure*}[t]
	%
	%
		\centering
		\includegraphics[width=1\linewidth]{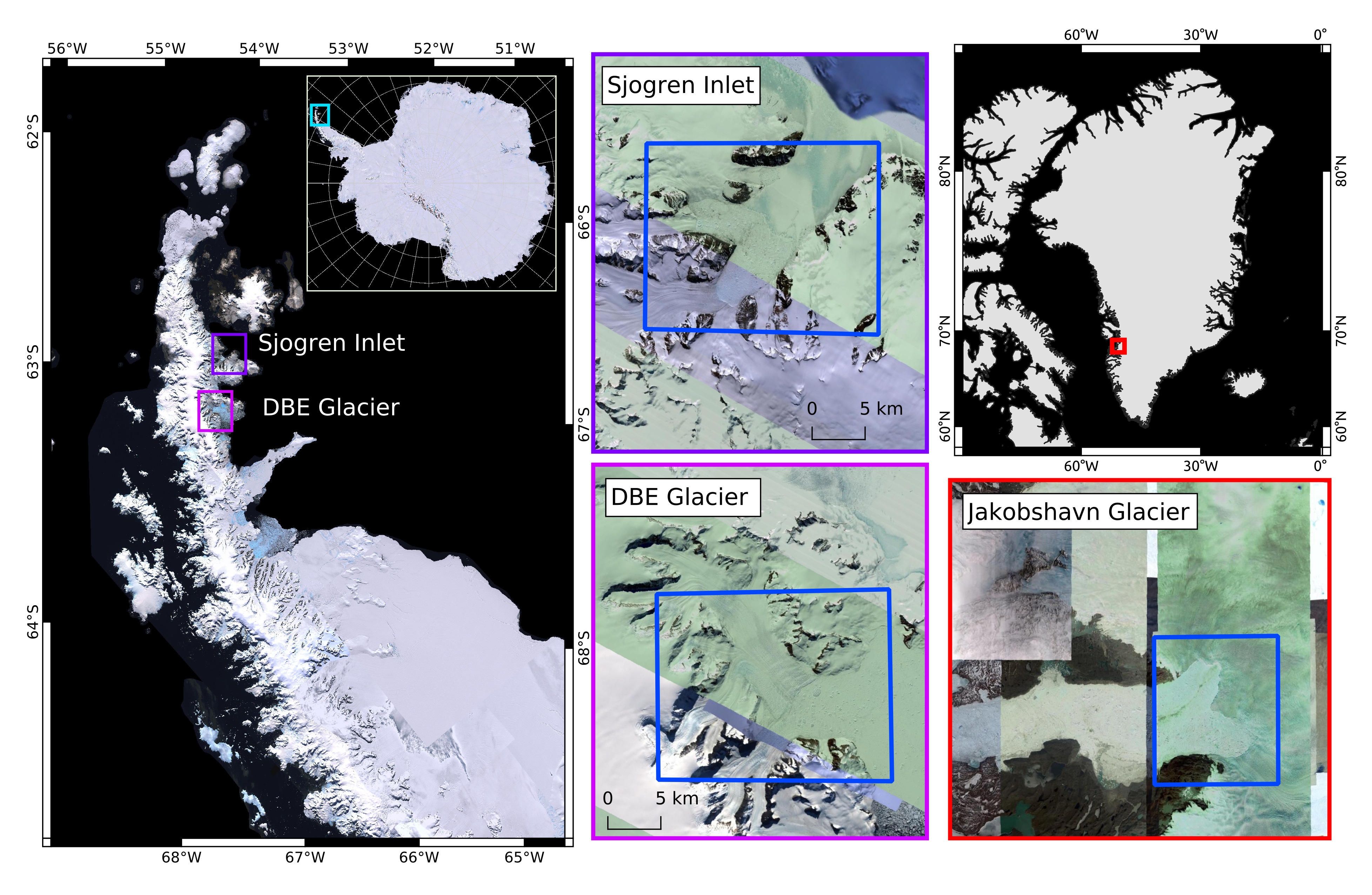}
	%
	\caption{Maps of the study regions and the locations of the studied glaciers at the Northern Antarctic Peninsula on the left, and Greenland Jakobshavn on the right. Blue polygons indicate subsets used for the CNN analysis. Left panels: Landsat LIMA Mosaic © USGS, NASA, BAS, NFS, Right panels: Bing Aerial Maps © Microsoft.
	}
	\label{fig:glaciers}
\end{figure*}

We perform our experiments on SAR images obtained from sites of the Antarctic Peninsula (AP) and Greenland, depicted in \cref{fig:glaciers}. AP has been strongly affected by changing climatic conditions~\cite{turner2016absence}. The Dinsmoor-Bombardier-Edgworth (DBE) and the Sjögren-Inlet (SI) glacier had undergone a strong frontal retreat over the years~\cite{seehaus2015changes,seehaus2016dynamic}. Jakobshavn Isbræ is one of the largest tidewater glaciers in Greenland and went through a drastic retreat over the last decades~\cite{joughin2008continued}.

The SAR images of the AP were calibrated and multi-looked to remove speckle noise. Furthermore, the images were geo-coded and ortho-rectified by means of the enhanced ASTER digital elevation model of the AP~\cite{cook2012new}. The ground truth of the front lines is taken from \etal{Seehaus}~\cite{seehaus2015changes, seehaus2016dynamic}. The images vary on the basis of their spatial resolution. High spatial resolution images cover an area of \SI{5x5}{\metre\square} per pixel and range until the lowest resolution of \SI{50x50}{\metre\square} per pixel. Out of 244 available SAR images at the AP, randomly 50 images were kept for testing, another 50 were used for validation, and the rest served for training the model. We increase the AP training dataset three times by augmenting with horizontally flipped image versions as well as by \SI{90}{\degree} rotated versions. The Jakobshavn Isbræ glacier calving front locations were provided by \etal{Zhang et al}~\cite{zhang2019automatically}. The SAR images of Jakobshavn Isbræ were preprocessed with a median filter to remove speckle noise. All the images have a constant spatial resolution of \SI{6x6}{\metre\square} per pixel. We use 25 random images for validation, 25 for testing, and the remaining 119 images as the training set. The training data is augmented by adding horizontally flipped image versions.

\begin{figure}[t]
	\centering
	\begin{subfigure}{0.5\linewidth}
		\centering
		\includegraphics[width=1\linewidth]{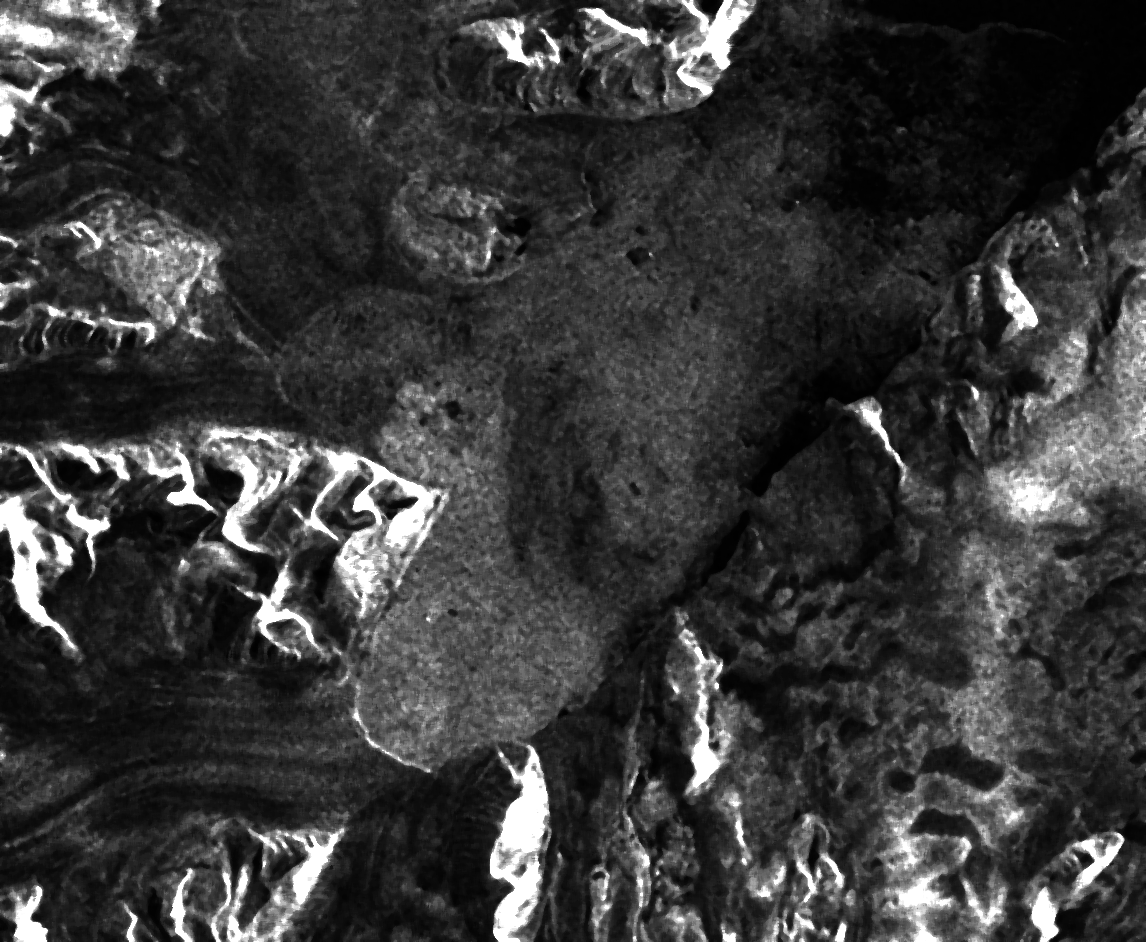}
		\caption{}
		\label{fig:gt_examples_a}
	\end{subfigure}
	
	\begin{subfigure}{0.48\linewidth}
		\centering
		\includegraphics[width=1\linewidth]{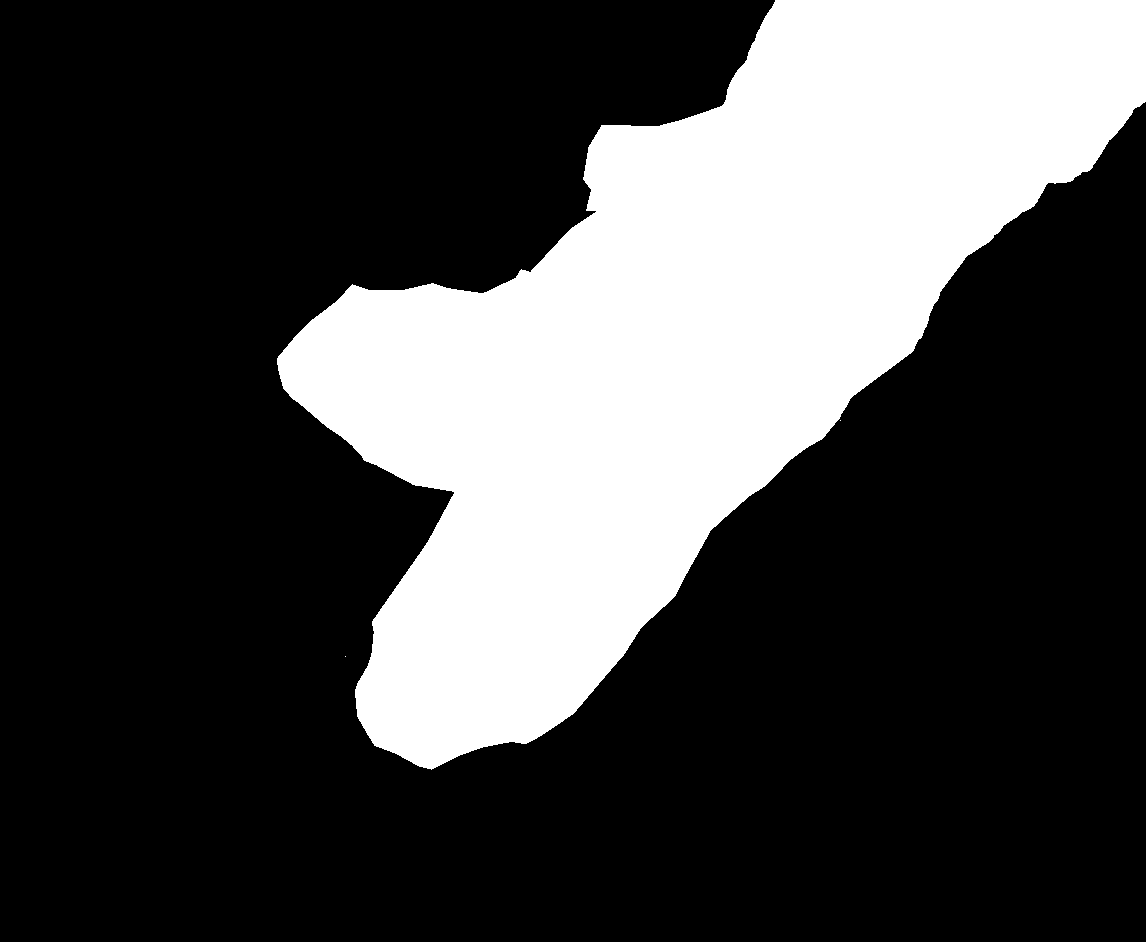}
		\caption{}
		\label{fig:gt_examples_b}
	\end{subfigure}
	\begin{subfigure}{0.48\linewidth}
		\centering
        \includegraphics[width=1\linewidth]{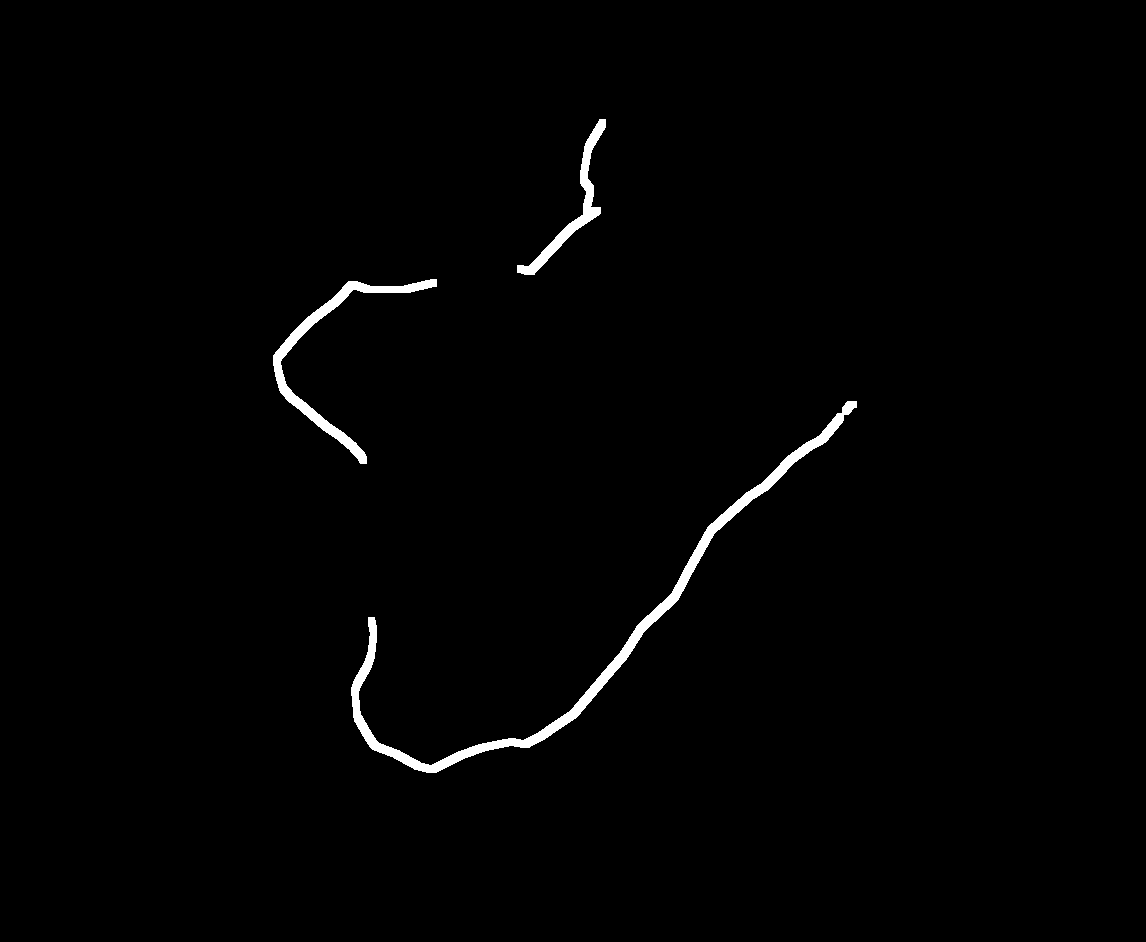}
        \caption{}
        \label{fig:gt_examples_c}
	\end{subfigure}
	\caption{\subref{fig:gt_examples_a} Example of a SAR image with its corresponding ground truth images: \subref{fig:gt_examples_b} ice/non-ice zones in which black represents the ice and white represents the non-ice zones, and \subref{fig:gt_examples_c} glacier calving front line ground truth. \\
	}
	\label{fig:gt_examples}
\end{figure}

The SAR images of AP also vary in size. The smallest of the images has dimensions of $377 \times 458$, and the largest one contains $3770 \times 4581$ pixels. Conversely, Jakobshavn Isbræ images are all approximately $1600\times3500$ pixels in size. For both datasets, two ground truth images are available: The calving front lines, and ice/non-ice zones. \Cref{fig:gt_examples} depicts a sample SAR image in our dataset, along with its corresponding ice/non-ice zones ground truth and its glacier calving front ground truth images.
The zones have a training class-imbalance of 1:3 in both datasets. The front line ground truth is severely imbalanced with a ratio of 1:2097 in AP training data and a ratio of 1:1124 in Jakobshavn Isbræ training data. 

\subsection{Experimental Setup}
\subsubsection{Architecture and Parameters}
We followed \etal{Zhang}~\cite{zhang2019automatically} as our baseline architecture. Both the convolution and transposed convolution layers have kernel sizes of $(5,5)$. We use leaky ReLUs as the activation functions with a negative slope of $\alpha=0.1$, and the pooling layers have a pooling size of $(2,2)$. For regularization, batch normalization layers have been exploited after every convolutional layer. Furthermore, we use Adam as the optimizer~\cite{kingma2014adam} and a cyclic learning rate for optimization, which is
beneficial to the learning of the network~\cite{smith2017cyclical}. Due to the constant changes in the network's performance by the cyclic training during training, we set a
rather high patience value (30 epochs) for early stopping.
Note that we also tried various other patience values for a subset of the experiments and the same behavior has been observed. Hence, we decided to use 30 for all experiments so that we give BCE early stopping criterion enough time to bring the validation loss down, too.
We use a patch size of $256\times 256$ pixels for patch extraction. The prediction of zones differs from the prediction of lines in the context of class-imbalance. Hence, we use slightly different parameters for these two different ground truths, which are explained in the following experiments. 

\subsubsection{Experiment 1: Early Stopping Criterion: MCC vs. BCE}
To demonstrate the importance of the metric used as the early stopping criterion, we perform the segmentation on (a) the zones and (b) on the lines, and compare the performance of BCE and MCC as the early stopping criteria. For segmenting zones, a batch size of $20$ is used. The maximum learning rate of $1e-2$ and a minimum of $1e-7$ is used in the cyclic learning rate with a step size of $5 \times$ epochs. To extract the calving front lines from the predicted zones, we post-process our predictions as follows: First, the false alarms are removed by extracting the largest connected component in the image. Then, we use canny edge detection~\cite{canny} to delineate the calving front lines from the rectified zones. 

In the second variant, where we train our model using the calving front lines directly, the front lines are one-pixel wide. This contributes to the major reason for the extreme class-imbalance. To facilitate the network's training on such a highly imbalanced dataset, the lines were thickened using morphological dilation with a rectangular structuring element of size $5 \times 5$ pixels. This reduces the class-imbalance of AP Dataset lines to $1:420$ and of Jakobshavn Isbræ to $1:225$. We use a batch size of $15$ for the front lines. The cyclic learning rates parameters for the lines are $1e-4$ and $1e-8$ as maximum and minimum, respectively.
 
\subsubsection{Experiment 2: Improved Distance Map BCE Loss vs.\ BCE loss}
After proving the fact that MCC as an early stopping criterion accomplishes better results, we perform our second experiment using MCC as an early stopping criterion. In the second experiment, we compare the performance of the improved distance map BCE loss (i.e., with optimum $k$) with the conventional BCE loss, inverse class-frequency weighted BCE, and the conventional distance map BCE loss (i.e., with $k=1$). This experiment is designed only to tackle high class-imbalance. Hence, the models are trained only using the front line ground truth images of both the datasets. We use the same parameters for predicting front line masks as in our first experiment. We search for the best parameters for the modified distance map formulation using grid search and thereby, select $w=3$ and $R=1$ for both datasets. The best value of $k$ was found to be $0.1$ for AP and $0.25$ for Jakobshavn Isbræ. 


\subsection{Evaluation}
\begin{figure}[t]
	\centering
\includegraphics[width=0.6\linewidth]{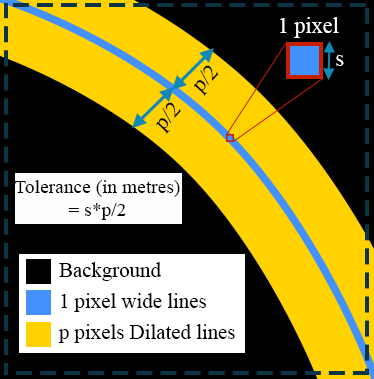}
	\caption{Tolerance scheme based on dilation of lines and the spatial resolution of image}
	\label{fig:toler}
\end{figure}

We use the performance metrics, that are commonly used for segmentation problems~\cite{taha2015metrics,minaee2020image}, \ie Intersection Over Union (IOU) and Dice coefficient, to evaluate our experiments. We also report the MCC. IOU is the area of overlap between the predicted masks $p$ and the ground truth masks $g$ divided by the area of union between the predicted masks and the ground truth masks. The Dice coefficient is twice the area of overlap divided by the total number of pixels from predictions and ground truth.
\begin{equation}
	\text{IOU} = \frac{p \cap g}{p \cup g}
\end{equation}
\begin{equation}
	\text{Dice} = \frac{2\sum\limits_{i=1}^{N} p_i g_i}{\sum\limits_{i=1}^{N}p_i + \sum\limits_{i=1}^{N}g_i}
\end{equation}
These performance metrics solely represent how well the predictions overlap with the ground truth masks. This makes it hard to evaluate the calving front lines because even if they are very close to the ground truth locations, they will be considered as a miss by these metrics. Therefore, to make these metrics more comprehensible, we further dilate the predictions and ground truth masks. The dilated pixels around the ground truth are considered as the relaxation or tolerance. Since our datasets also come with spatial resolution, we calculate the tolerance in meters. The spatial resolution $s$ denotes the actual distance one pixel covers on earth. Therefore, the tolerance $t$ is calculated as $t = s \times p/2$. The tolerance scheme is depicted in \cref{fig:toler}. The metrics are reported in our tables along with the tolerance (in meters) for the predicted calving front lines.

\begin{figure}[tbhp]
	\begin{subfigure}[b]{0.325\linewidth}
		\centering
		\includegraphics[width=1\linewidth]{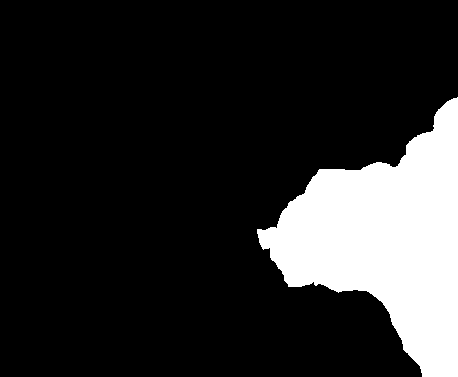}
		
		\vspace{3px}
		\includegraphics[width=1\linewidth]{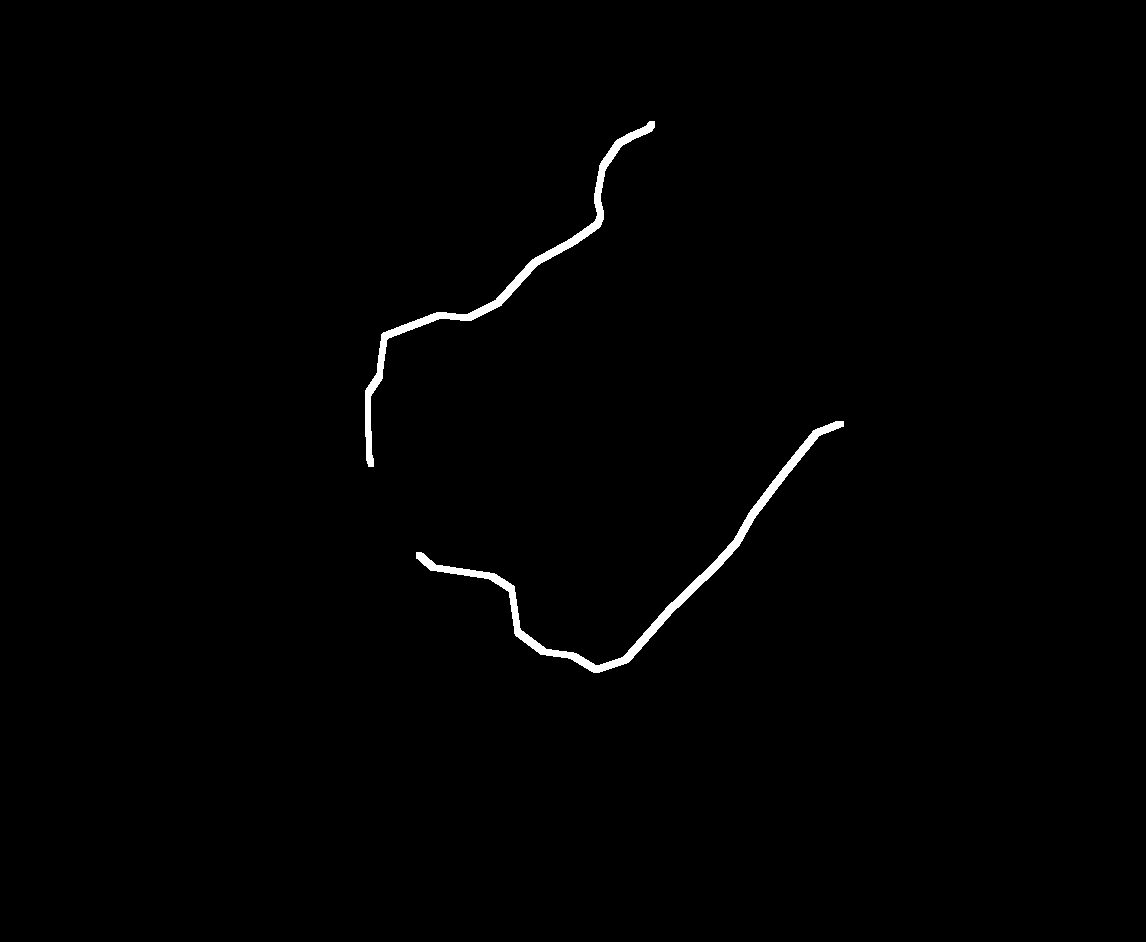}
		\caption{}	\label{fig:amccmonitor_a}
	\end{subfigure}
	\begin{subfigure}[b]{0.325\linewidth}
		\centering
		\includegraphics[width=1\linewidth]{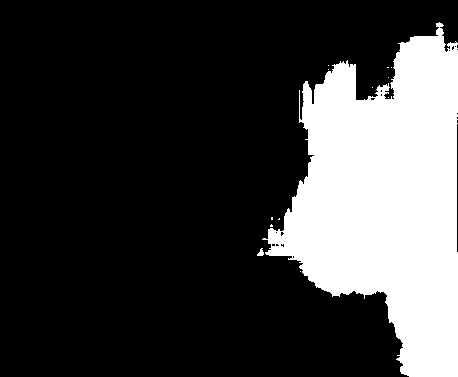}
		
		\vspace{3px}
		\includegraphics[width=1\linewidth]{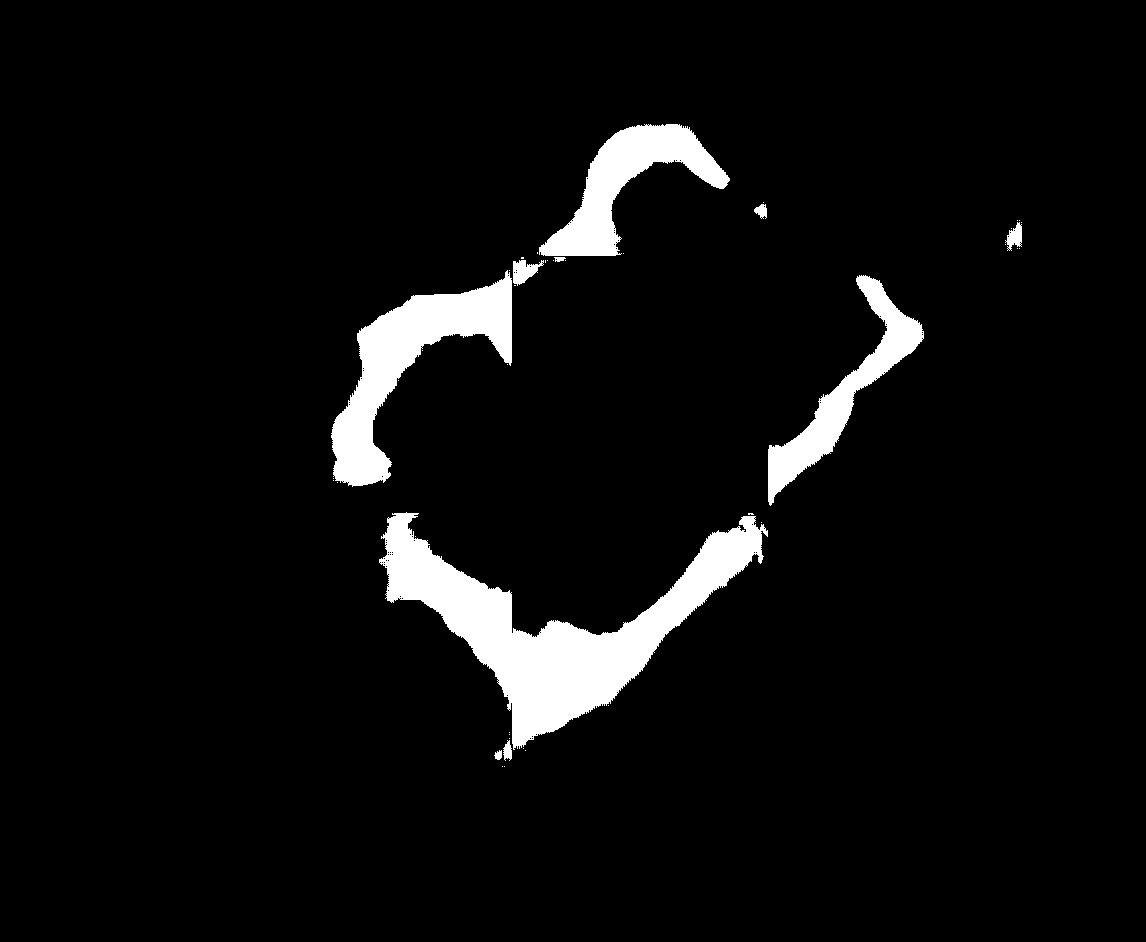}
		\caption{}	\label{fig:amccmonitor_b}
	\end{subfigure}
	\begin{subfigure}[b]{0.325\linewidth}
		\centering
		\includegraphics[width=1\linewidth]{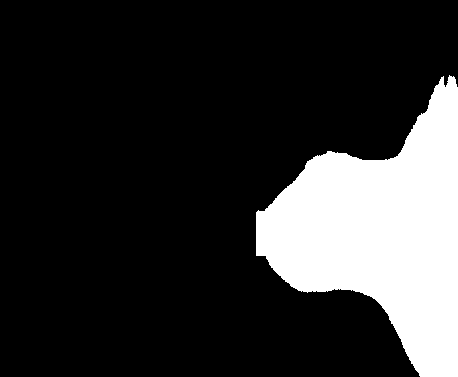}
		
		\vspace{3px}
		\includegraphics[width=1\linewidth]{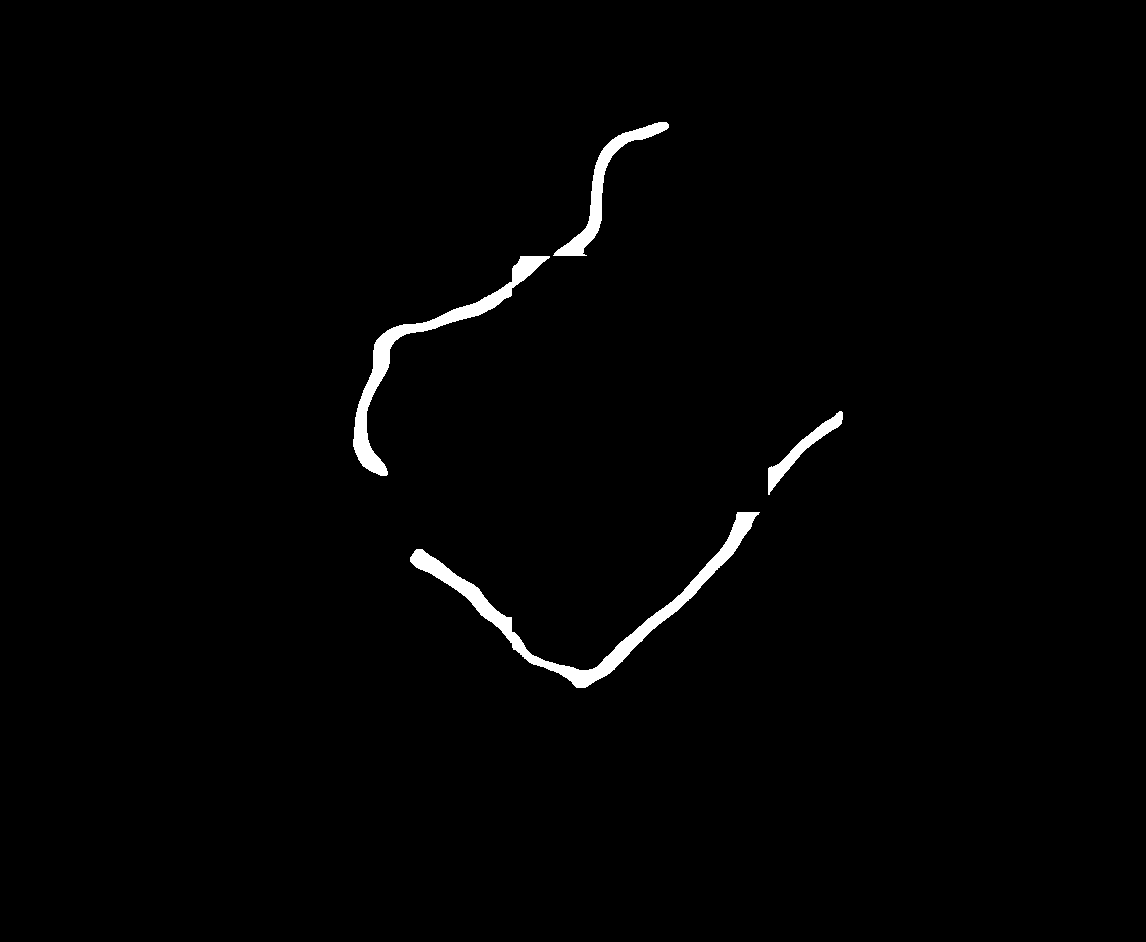}
		\caption{}	\label{fig:amccmonitor_c}
	\end{subfigure}
	\caption{Comparing the segmentation results using BCE and MCC as the early stopping criterion on AP glaciers: \subref{fig:amccmonitor_a} ground truth, \subref{fig:amccmonitor_b} segmentation maps predicted using BCE as the early stopping criterion, \subref{fig:amccmonitor_c} segmentation maps predicted using MCC as the early stopping criterion.
	}
	\label{fig:amccmonitor}
\end{figure}

\begin{figure}[tbhp]
	\begin{subfigure}{1\linewidth}
		\centering
		\includegraphics[width=1\linewidth]{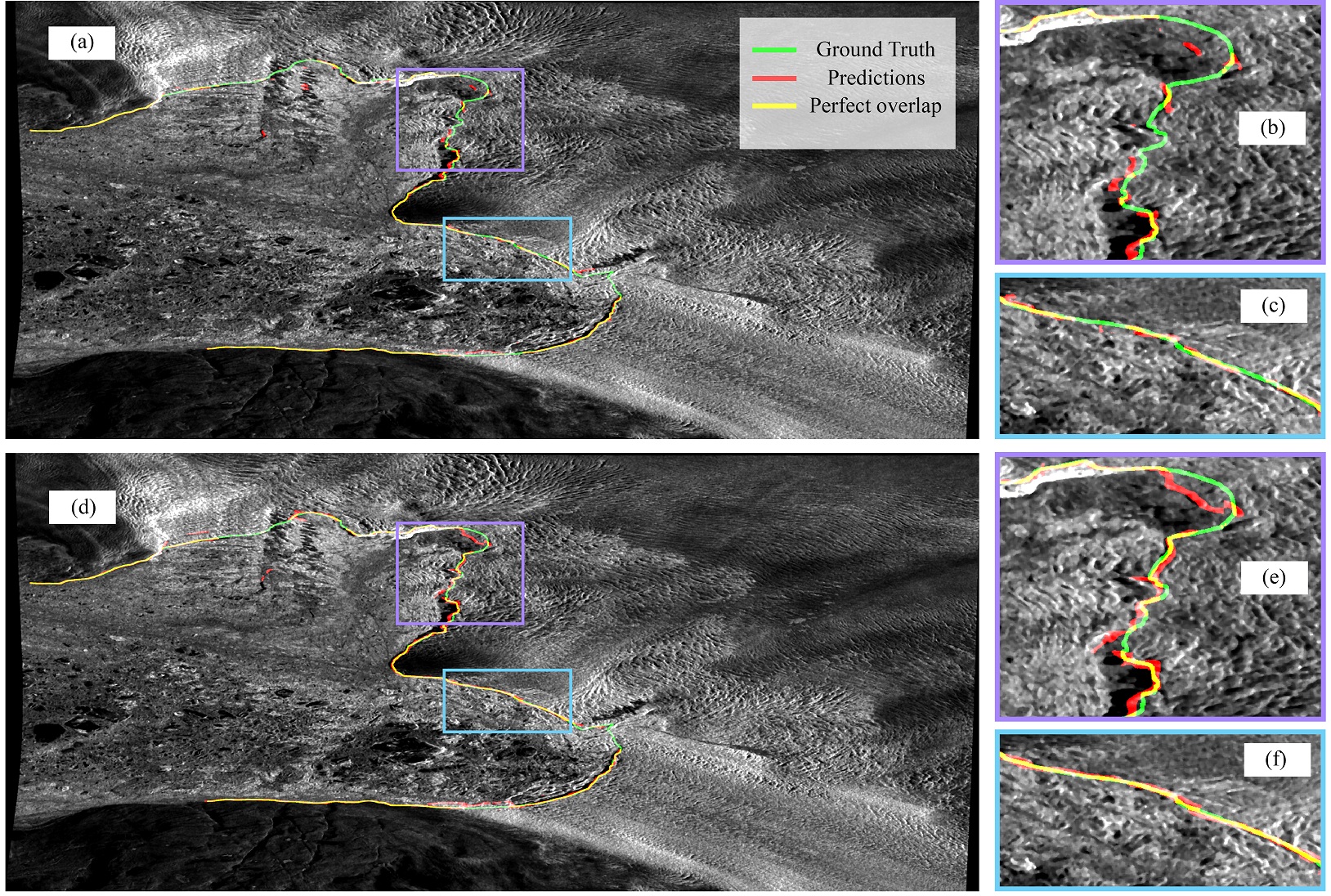}
		\phantomsubcaption\label{fig:qualdbce_a}%
		\phantomsubcaption\label{fig:qualdbce_b}%
		\phantomsubcaption\label{fig:qualdbce_c}%
		\phantomsubcaption\label{fig:qualdbce_d}%
		\phantomsubcaption\label{fig:qualdbce_e}%
		\phantomsubcaption\label{fig:qualdbce_f}%
	\end{subfigure}
		%
	\caption{Comparison of predictions of \subref{fig:qualdbce_a} BCE Loss, and \subref{fig:qualdbce_d} DMap BCE Loss. \subref{fig:qualdbce_b} and \subref{fig:qualdbce_c} show magnified versions of the marked areas from the BCE Loss prediction image, and \subref{fig:qualdbce_e} and \subref{fig:qualdbce_f} show magnified areas from the DMap BCE loss prediction image.}
	\label{fig:qualdbce}
\end{figure}

\begin{figure}[tbhp]
	\begin{subfigure}{0.32\linewidth}
		\centering
		\includegraphics[width=1\linewidth]{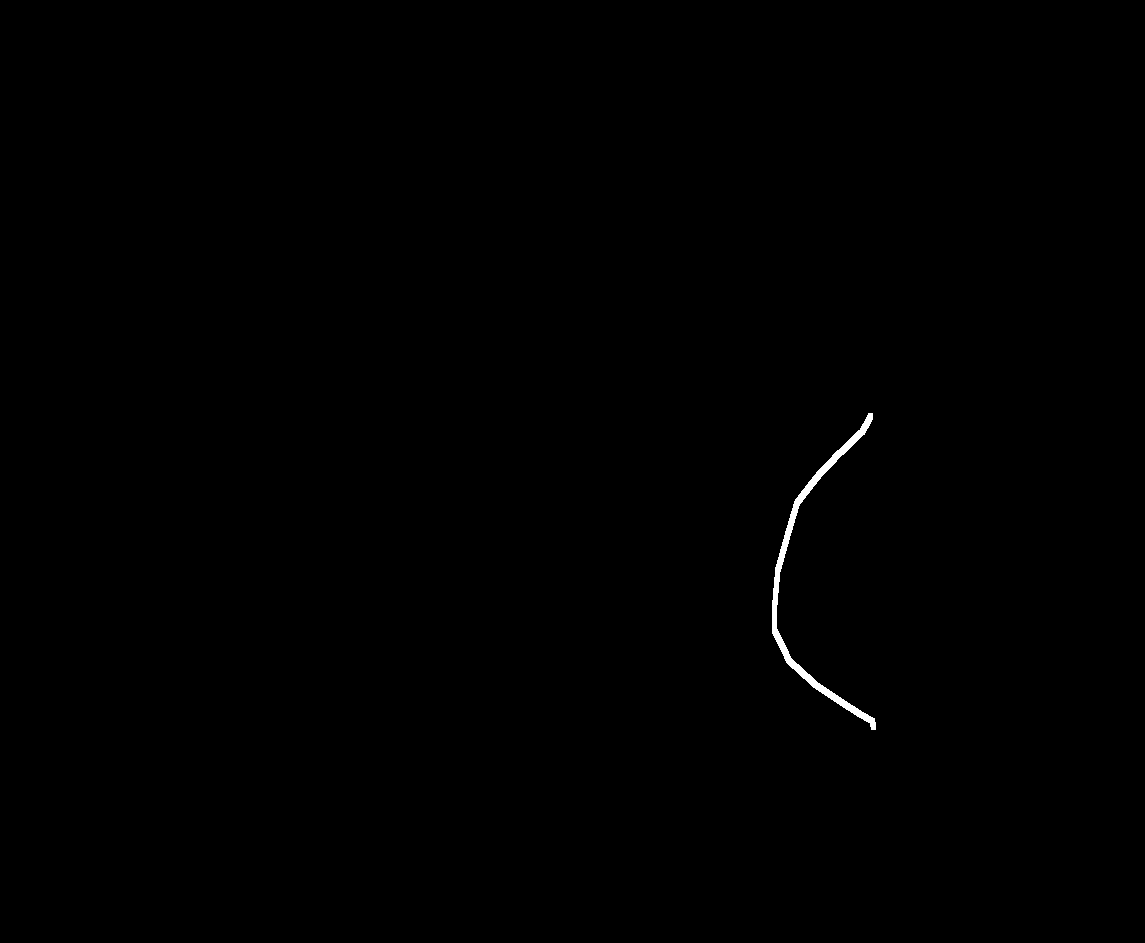}
		
		\vspace{3px}
		\includegraphics[width=1\linewidth]{Figures/1lines_gtm_thickened.png}
		\caption{}\label{fig:dbce_a}
	\end{subfigure}
	\begin{subfigure}{0.32\linewidth}
		\centering
	\includegraphics[width=1\linewidth]{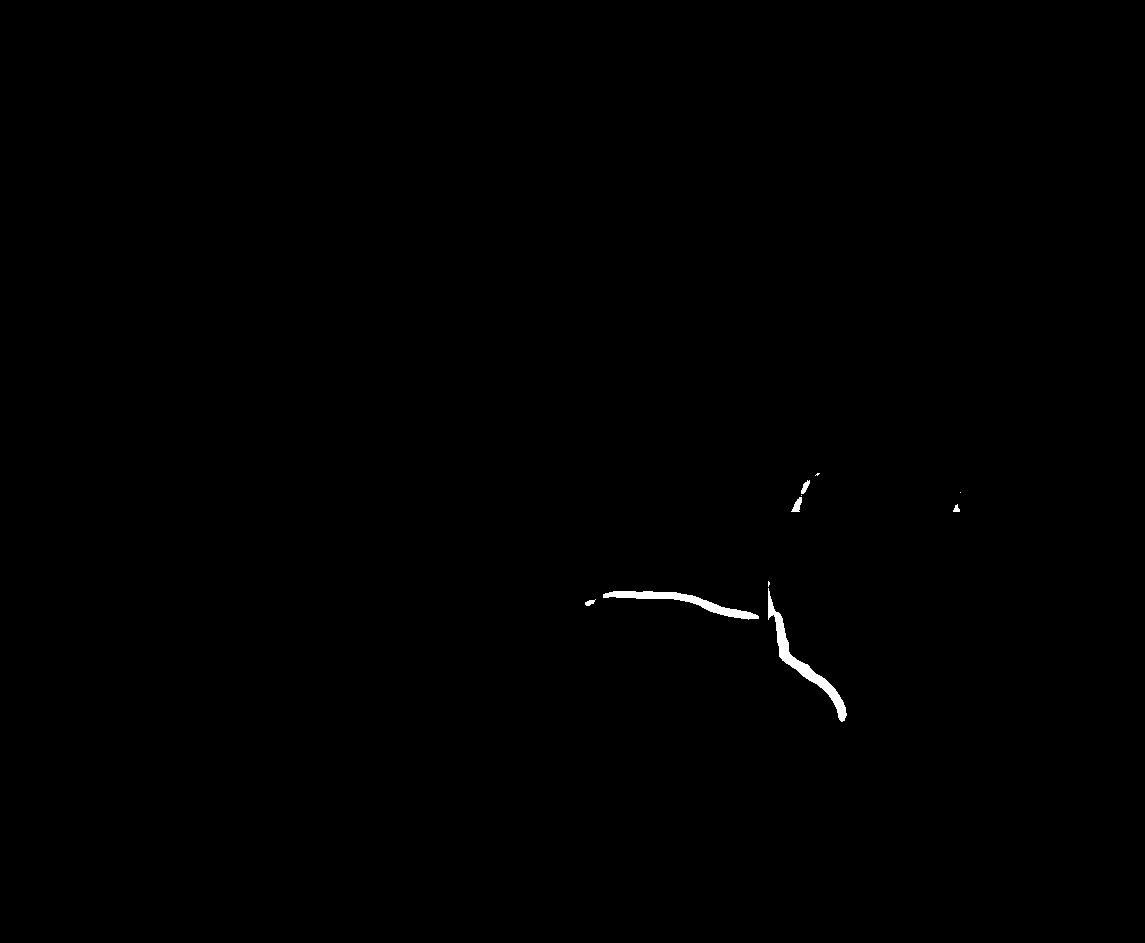}
		
		\vspace{3px}
		\includegraphics[width=0.94\linewidth]{Figures/1lines_mmcc.png}
		\caption{}\label{fig:dbce_b}
	\end{subfigure}
	\begin{subfigure}{0.32\linewidth}
		\centering
\includegraphics[width=1\linewidth]{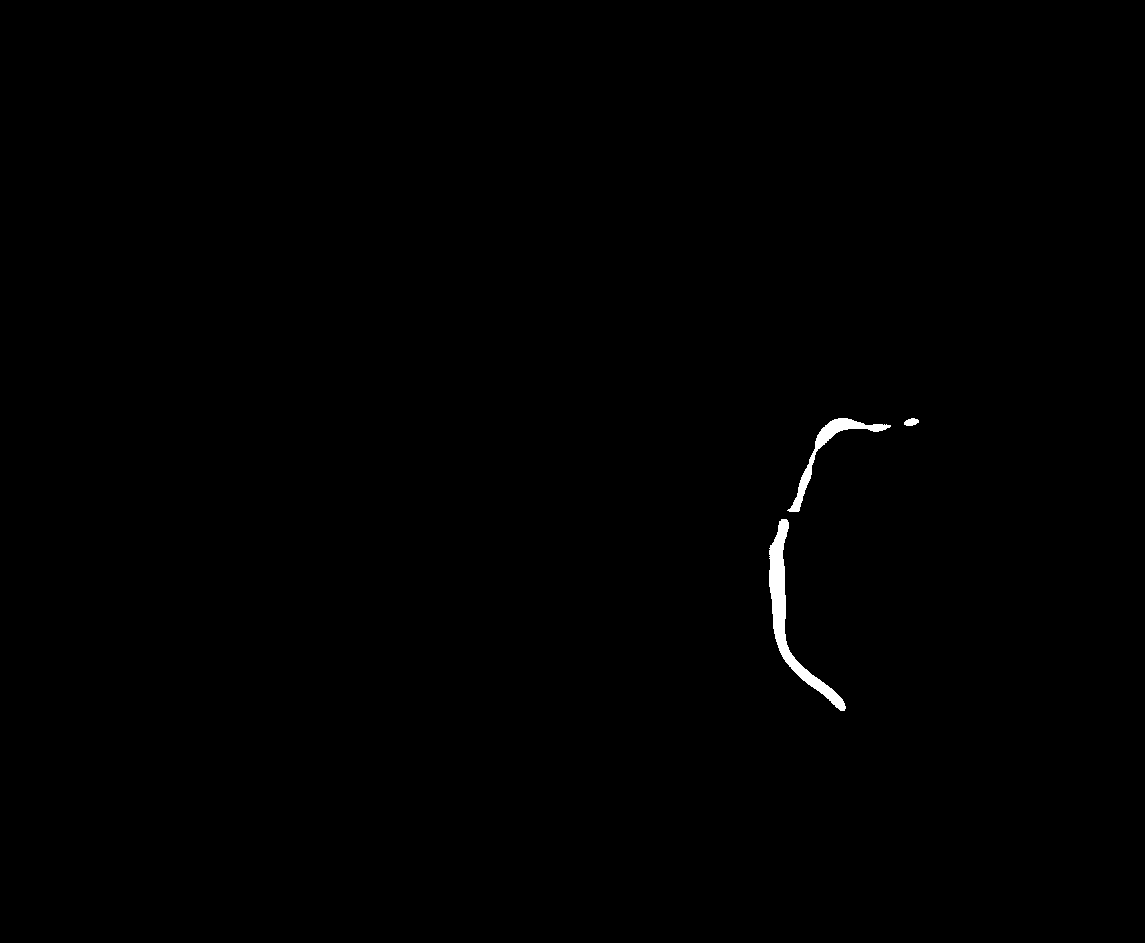}
		
		\vspace{3px}
		\includegraphics[width=1\linewidth]{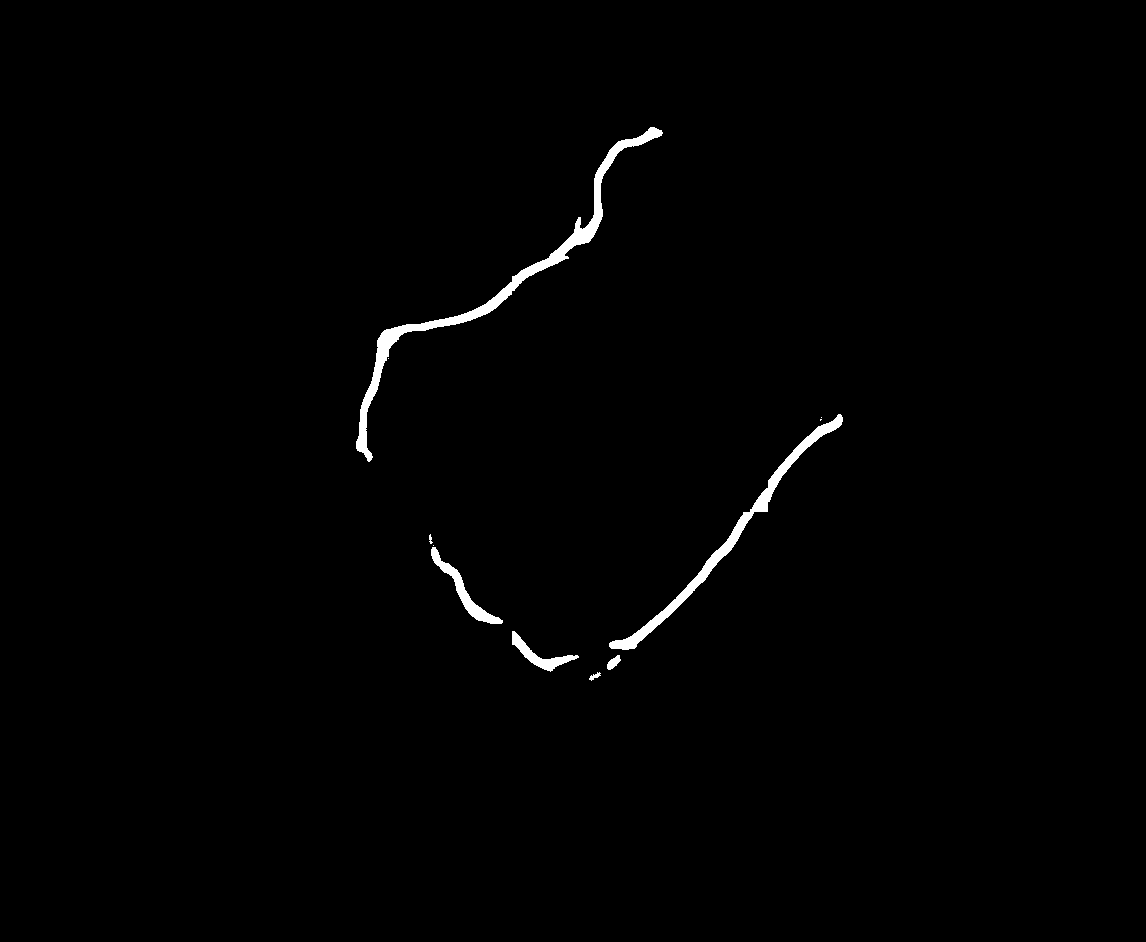}
		\caption{}\label{fig:dbce_c}
	\end{subfigure}
	\caption{Predicted lines achieved by \subref{fig:dbce_b} BCE loss, and \subref{fig:dbce_b} distance map BCE loss. Image \subref{fig:dbce_a} depicts the ground truth. The two samples are from the AP glaciers dataset.}
	\label{fig:dbce}
\end{figure}

\Cref{tab:EstopMCC} shows the results when using BCE and MCC as an early stopping criterion. Monitoring MCC clearly outperforms BCE in all cases. Prediction of zones in Jakobshavn Isbræ does not show much improvement as BCE as early stopping performed very well, having IOU and Dice score of around \SI{97}{\percent}. 
The reason is that the use of MCC continues training to roughly 150 epochs, while BCE loss already stops at around 50 epochs. 
Validation BCE decides that the network is overfitting too early. 
On the other hand, MCC is more comprehensive about the validation error and correctly allows the network to be trained further. \Cref{tab:EstopMCC} also illustrates that monitoring MCC shows improvement not only when dealing with high class-imbalance, but also is quite reliable for less imbalanced datasets (AP and Jakobshavn Isbræ zones).
\begin{table}[tbp]
	\centering
	\caption{Performance when using BCE and MCC as early stopping criterion.}
	\begin{tabular}{ccp{4.355em}cc}
	\toprule
		\multicolumn{1}{p{5.785em}}{Datasets} & \multicolumn{1}{p{4.785em}}{Metrics} & \multicolumn{1}{p{4.355em}|}{Tolerance} & \multicolumn{2}{p{10.36em}}{Early Stopping Criterion} \\
		\cmidrule{1-3}          &       & \multicolumn{1}{c|}{} & BCE   & MCC \\
		\cmidrule{4-5}    \multicolumn{1}{p{5.785em}}{AP} & \multicolumn{1}{p{4.785em}}{IOU} & \multicolumn{1}{c}{} & 0.8264 & \textbf{0.8912} \\
		\cmidrule{2-5}    \multicolumn{1}{p{5.785em}}{zones} & \multicolumn{1}{p{4.785em}}{Dice Coeff} & \multicolumn{1}{c}{} & 0.8984 & \textbf{0.9389} \\
		\cmidrule{2-5}          & \multicolumn{1}{p{4.785em}}{MCC} & \multicolumn{1}{c}{} & 0.8764 & \textbf{0.9254} \\
		\midrule
		& \multicolumn{1}{l}{\multirow{3}[2]{*}{IOU}} & 65 m  & 0.1524 & \textbf{0.2817} \\
		&       & 110 m & 0.2201 & \textbf{0.3957} \\
		&       & 155 m & 0.2712 & \textbf{0.4561} \\
		\cmidrule{2-5}          & \multicolumn{1}{l}{\multirow{3}[2]{*}{Dice Coeff}} & 65 m  & 0.2617 & \textbf{0.4421} \\
		\multicolumn{1}{p{5.785em}}{AP} &       & 110 m & 0.3549 & \textbf{0.5385} \\
		\multicolumn{1}{p{5.785em}}{lines} &       & 155 m & 0.4196 & \textbf{0.5997} \\
		\cmidrule{2-5}          & \multicolumn{1}{l}{\multirow{3}[2]{*}{MCC}} & 65 m  & 0.3328 & \textbf{0.4496} \\
		&       & 110 m & 0.4112 & \textbf{0.5456} \\
		&       & 155 m & 0.4657 & \textbf{0.6062} \\
		\midrule
		\multicolumn{1}{p{5.785em}}{Jakobshavn} & \multicolumn{1}{p{4.785em}}{IOU} & \multicolumn{1}{c}{} & 0.9613 & \textbf{0.9666} \\
		\cmidrule{2-5}    \multicolumn{1}{p{5.785em}}{Isbræ} & \multicolumn{1}{p{4.785em}}{Dice Coeff} & \multicolumn{1}{c}{} & 0.9801 & \textbf{0.9828} \\
		\cmidrule{2-5}    \multicolumn{1}{p{5.785em}}{zones} & \multicolumn{1}{p{4.785em}}{MCC} & \multicolumn{1}{c}{} & 0.9725 & \textbf{0.9763} \\
		\midrule
		& \multicolumn{1}{l}{\multirow{3}[2]{*}{IOU}} & 60 m  & 0.3755 & \textbf{0.5681} \\
		&       & 105 m & 0.4424 & \textbf{0.6451} \\
		&       & 150 m & 0.4801 & \textbf{0.6928} \\
		\cmidrule{2-5}    \multicolumn{1}{p{5.785em}}{Jakobshavn} & \multicolumn{1}{l}{\multirow{3}[2]{*}{Dice Coeff}} & 60 m  & 0.5437 & \textbf{0.7223} \\
		\multicolumn{1}{p{5.785em}}{Isbræ} &       & 105 m & 0.6108 & \textbf{0.7824} \\
		\multicolumn{1}{p{5.785em}}{lines} &       & 150 m & 0.6459 & \textbf{0.8167} \\
		\cmidrule{2-5}          & \multicolumn{1}{l}{\multirow{3}[2]{*}{MCC}} & 60 m  & 0.5712 & \textbf{0.7208} \\
		&       & 105 m & 0.6273 & \textbf{0.7783} \\
		&       & 150 m & 0.6553 & \textbf{0.8103} \\
		\bottomrule
	\end{tabular}%
	\label{tab:EstopMCC}%
\end{table}%
%
%
The fact that the BCE-monitored network is not trained enough can be clearly observed in the qualitative results that are depicted in \cref{fig:amccmonitor}. 
The MCC-monitored network performs significantly better for both zones and lines. Apart from the fact that the calving front lines generated from the MCC monitored network are more accurate, the predictions also seem to have fewer numbers of false positives (incorrect white lines on the black regions). 
The results of BCE Loss and improved distance map BCE Loss are shown in \cref{tab:dmbceloss}. The improved distance map BCE loss is also compared with the distance map using $(k=1)$. It is interesting to notice that the distance map BCE loss with $k=1$ did not improve the performance of BCE loss in our task of glacier front line segmentation. However, it does show slight improvement with low tolerance (\SI{65}{m}) for the AP glaciers. In contrast, the optimum $k$ helps to deal with the class-imbalance better. The qualitative results of the predicted lines on Jakobshavn Isbræ regions are depicted in \cref{fig:qualdbce}. In this figure, the green color represents ground truth, the red color shows the incorrect prediction, and the yellow color depicts the correct predictions, \ie the overlap between the prediction and the ground truth. The line predictions of Jakobshavn Isbræ glaciers in both losses are disconnected. However, the distance map loss manages to identify more regions of the calving fronts. BCE loss also incorrectly predicts severe outliers (false positives) that are far away from the real locations as depicted in \cref{fig:qualdbce}. Such false positives are less numerous in the case of the distance map BCE predictions.
\begin{table*}[tb]
	\centering
	\caption{Performance of the improved distance map BCE Loss (\ie with optimum $k$) vs.\ the BCE loss, inverse class frequency weighted BCE, and the conventional distance map BCE Loss.}
	\begin{tabular}{llccccc}
	\toprule
		Datasets & Metrics & \multicolumn{1}{c|}{Tolerance (meters)} & \multicolumn{4}{c}{Loss function} \\
		\cmidrule{1-3}          &       & \multicolumn{1}{c|}{} & BCE   & \multicolumn{1}{p{4.43em}}{BCE (with weights)} & \multicolumn{1}{p{5.5em}}{Dmap BCE, $k=1$} & \multicolumn{1}{p{5.145em}}{Dmap BCE, optimum $k$} \\
		\cmidrule{4-7}    \multicolumn{1}{l}{\multirow{9}[6]{*}{AP lines}} & \multicolumn{1}{l}{\multirow{3}[2]{*}{IOU}} & 65  & 0.2817 & 0.2798 & 0.3078 & \textbf{0.3085} \\
		&       & 110 & 0.3957 & 0.3728 & 0.3991 & \textbf{0.4072} \\
		&       & 155 & 0.4561 & 0.4355 & 0.4591 & \textbf{0.4734} \\
		\cmidrule{2-7}          & \multicolumn{1}{l}{\multirow{3}[2]{*}{Dice Coeff.}} & 65  & 0.4421 & 0.4211 & 0.4456 & \textbf{0.4463} \\
		&       & 110 & 0.5385 & 0.5242 & 0.5442 & \textbf{0.5506} \\
		&       & 155 & 0.5997 & 0.5873 & 0.6046 & \textbf{0.6152} \\
		\cmidrule{2-7}          & \multicolumn{1}{l}{\multirow{3}[2]{*}{MCC}} & 65  & 0.4496 & 0.4302 & 0.4502 & \textbf{0.4544} \\
		&       & 110 & 0.5456 & 0.5295 & 0.5467 & \textbf{0.5566} \\
		&       & 155 & 0.6062 & 0.5913 & 0.6059 & \textbf{0.6204} \\
		\midrule
		\multicolumn{1}{l}{\multirow{9}[6]{*}{Jakobshavn Isbræ lines}} & \multicolumn{1}{l}{\multirow{3}[2]{*}{IOU}} & 60  & 0.5681 & 0.5535 & 0.5675 & \textbf{0.5908} \\
		&       & 105 & 0.6451 & 0.6272 & 0.6417 & \textbf{0.6714} \\
		&       & 150 & 0.6928 & 0.6717 & 0.6872 & \textbf{0.7182} \\
		\cmidrule{2-7}          & \multicolumn{1}{l}{\multirow{3}[2]{*}{Dice Coeff.}} & 60  & 0.7223 & 0.7101 & 0.7217 & \textbf{0.7401} \\
		&       & 105 & 0.7824 & 0.7684 & 0.7794 & \textbf{0.801} \\
		&       & 150 & 0.8167 & 0.8012 & 0.8123 & \textbf{0.8338} \\
		\cmidrule{2-7}          & \multicolumn{1}{l}{\multirow{3}[2]{*}{MCC}} & 60  & 0.7208 & 0.707 & 0.7204 & \textbf{0.7364} \\
		&       & 105 & 0.7783 & 0.7031 & 0.7754 & \textbf{0.7957} \\
		&       & 150 & 0.8103 & 0.7937 & 0.8058 & \textbf{0.8271} \\
		\bottomrule
	\end{tabular}%
	\label{tab:dmbceloss}%
\end{table*}%
Referring to \cref{fig:dbce}, which depicts the front line predictions of the AP glaciers, we see that the BCE loss predicts several images quite inaccurately. These images are well handled by the distance map loss.

\begin{figure}[tbhp]
	\begin{subfigure}[b]{0.48\linewidth}
		\centering
		\includegraphics[width=1\linewidth]{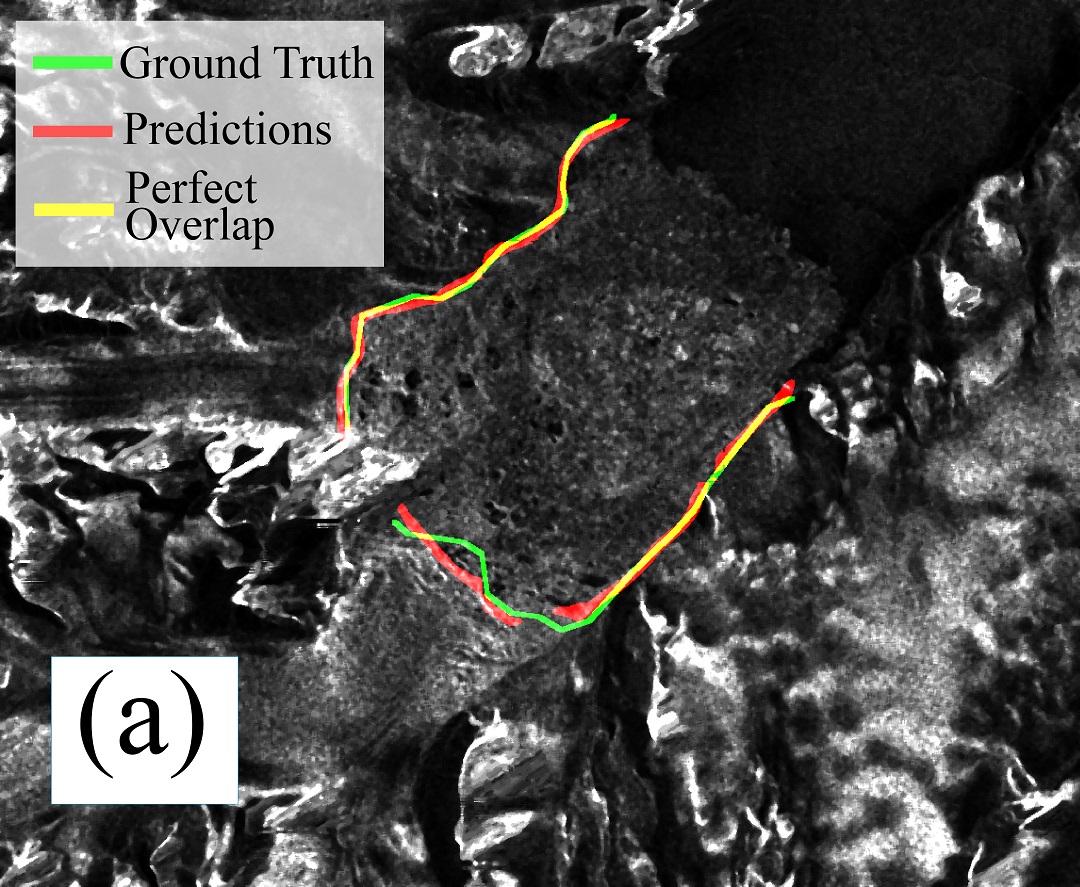}
	\caption{}\label{fig:1qual_a}
	\end{subfigure}
	\begin{subfigure}[b]{0.48\linewidth}
		\centering
	\includegraphics[width=1\linewidth]{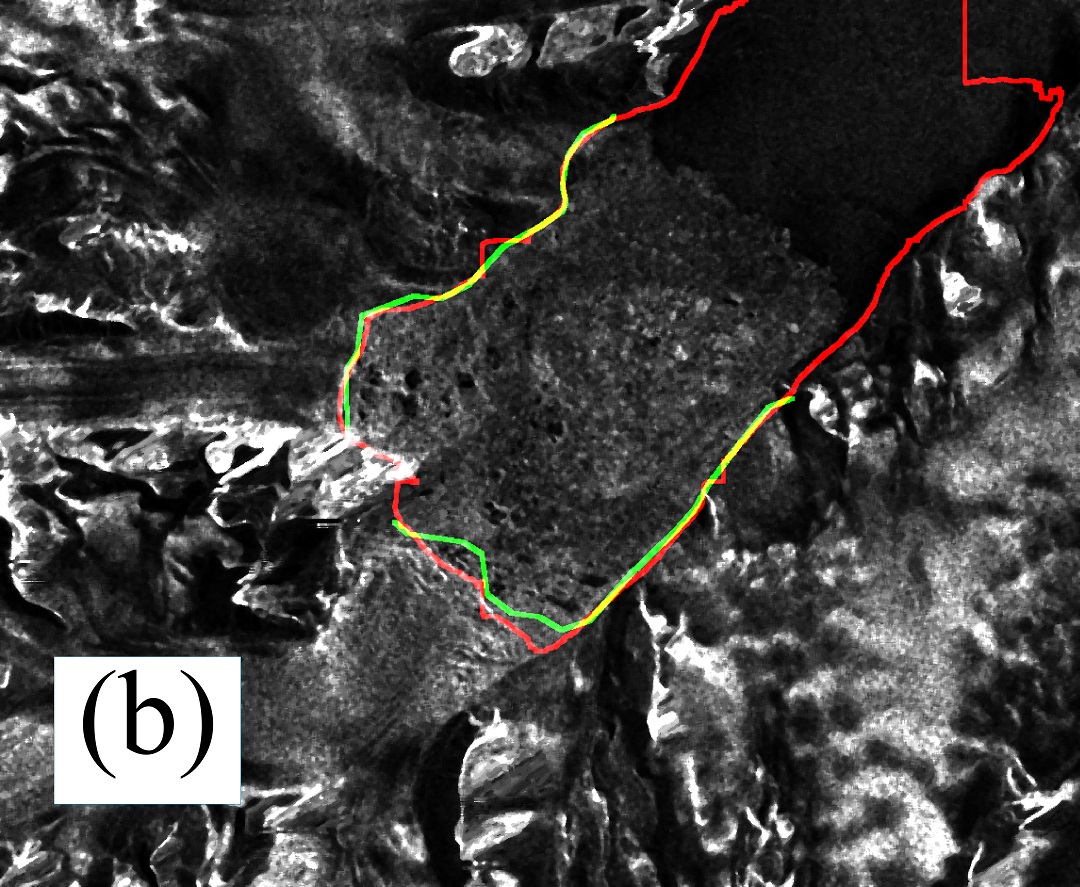}
		\caption{}\label{fig:1qual_b}
	\end{subfigure}
	\caption{Comparison of the predictions achieved by models trained on \subref{fig:1qual_a} front lines, and  \subref{fig:1qual_b} zones and then post-processed. The sample is from the AP glaciers dataset.}
	\label{fig:1qual}
\end{figure}

\begin{figure}[tbhp]
		\includegraphics[width=1\linewidth]{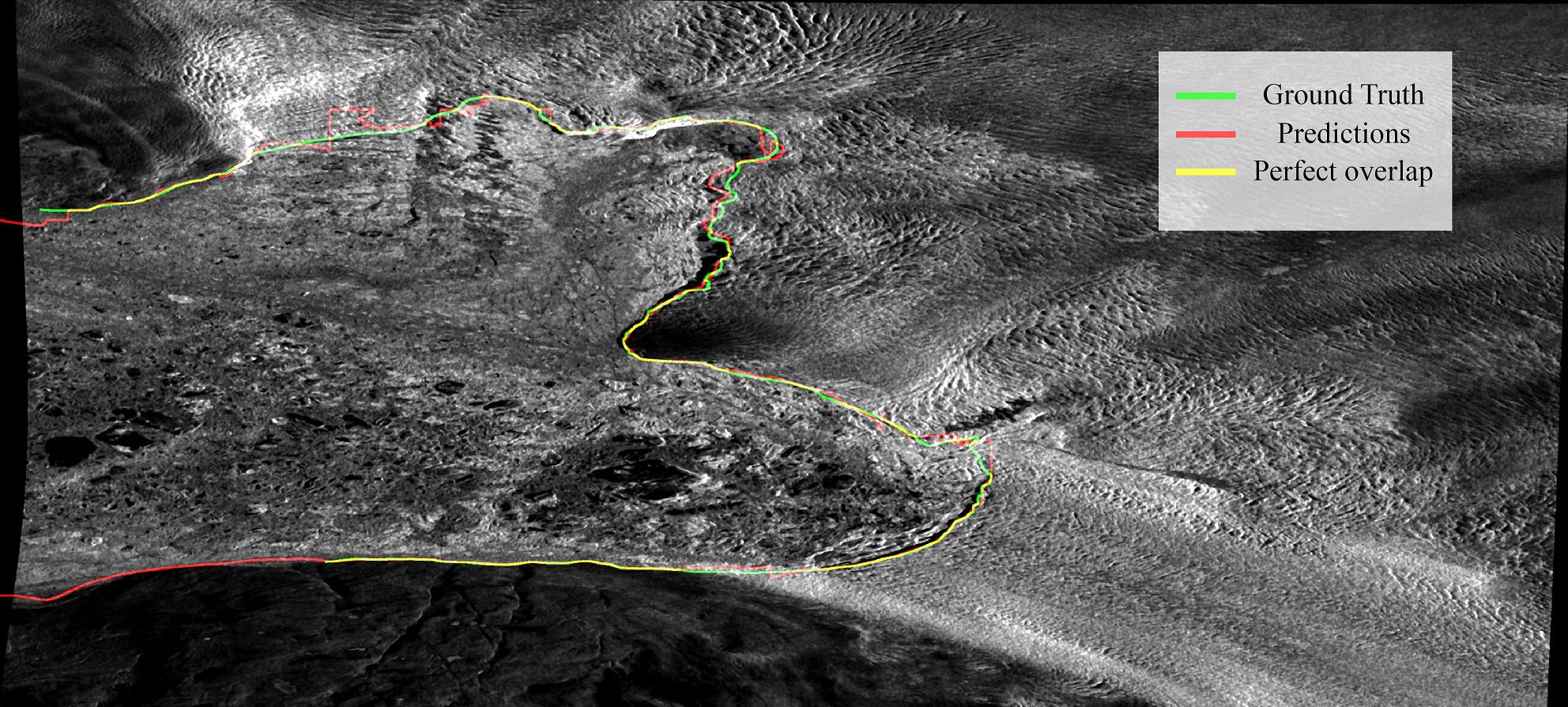}
		%
	\caption{Calving front prediction of Jakobshavn Isbræ computed by postprocessing the predicted ice/non-ice zones.}
	\label{fig:3qualzones}
\end{figure}

When working on the problem of automatic detection of CFPs in the SAR images, the available ground truth data for the training could be the ice zones or the calving front lines. The semantic segmentation of zones on the SAR images needs post-processing to find the calving front locations. Unlike semantically segmenting the zones, a direct semantic segmentation of the calving front locations is a harder problem due to the more severe class-imbalance. \Cref{fig:1qual} qualitatively compares the segmented front lines using the two approaches for the AP glaciers dataset: one approach predicts the front lines directly and the other post-processes the predicted zones. Direct predictions of calving front lines bring minor disjoints in the predictions, whereas the post-processed zone predictions are continuous. However, the result of the post-processed zone is not entirely the glacier calving front lines. It shows the boundary between ice zones and non-ice zones, which may contain some bedrock boundaries as well. This is also observable on the post-processed predictions for the Jakobshavn glacier in \cref{fig:3qualzones}. Moreover, some regions are better predicted by direct front line predictions.
%


\section{Conclusion}\label{sec:conclusion}
This paper proposes methods to improve deep learning-based semantic segmentation of the glacier calving fronts. The detection of calving fronts is challenged by its severe class inequality. We have tackled the class-imbalance problem by two approaches. First, the use of Mathews Correlation Coefficient as an early stopping criteria showed a great performance gain. Then, to help the binary cross-entropy loss function to handle the imbalanced data, we incorporated a novel distance map loss. 
The distance map can be used in conjunction with any other loss function. The MCC-monitored network outperforms the BCE-monitored network by about \SI{15}{\percent} dice coefficient and IOU. Using the modified distance map BCE loss results in a further improvement of approximately \SI{2}{\percent} dice coefficient. 

Finally, two approaches of predicting front lines were compared, \ie either to directly predict the front lines or to predict ice zones and then post-process the areas to delineate the calving fronts. The presented results in this work are encouraging and we believe that other image segmentation applications with severe-class-imbalance can benefit from the proposed approaches in this work.
\bibliographystyle{ieeetr}
\bibliography{References}

\end{document}